\documentclass[twocolumn,showpacs,prb,citeautoscript,amsmath,amssymb,floatfix,eqsecnum]{revtex4-2}
\usepackage{amsmath,amssymb,color}
\usepackage{bm}
\usepackage{graphicx}
\usepackage{psfrag}
\usepackage{bbold}
\usepackage{bbm}
\usepackage{soul}
\usepackage{filecontents}
\usepackage{babel}

\newcommand{\bd}{\bm}

\begin{document}
	
	\title{Collective modes in the charge-density-wave state of K$_{0.3}$MoO$_3$: The role of long-range Coulomb interactions revisited
	}


	\author{Max O. Hansen$^1$, Yash Palan$^1$,
		Viktor Hahn$^1$, Mark D. Thomson$^2$, Konstantin Warawa$^2$, Hartmut G. Roskos$^2$,  Jure Demsar$^3$, Falko Pientka$^1$, Oleksandr Tsyplyatyev$^1$, and Peter Kopietz$^1$}
	
	\affiliation{$^1$Institut f\"{u}r Theoretische Physik, Universit\"{a}t
		Frankfurt,  Max-von-Laue Strasse 1, 60438 Frankfurt, Germany}
	
	\affiliation{$^2$Physikalisches Institut, Universit\"{a}t
		Frankfurt,  Max-von-Laue Strasse 1, 60438 Frankfurt, Germany}
	
	\affiliation{$^3$Institut f\"{u}r Physik, Universit\"{a}t Mainz, Staudingerweg 7, 55128 Mainz, Germany}
	
	\date{March 15, 2023}

	\begin{abstract}
		We re-examine the effect of long-range Coulomb interactions on the collective amplitude and phase modes in the incommensurate charge-density-wave ground state of quasi-one-dimensional conductors.	Using an effective action approach  we show that the longitudinal acoustic phonon protects the gapless linear dispersion of the lowest phase mode in the presence of long-range Coulomb interactions. Moreover, in Gaussian approximation, amplitude fluctuations are not affected by long-range Coulomb interactions. We also calculate the collective mode dispersions at finite temperatures and compare our results with the measured energies of amplitude and phase modes in K$_{0.3}$MoO$_3$. With the exception of the lowest phase mode, the temperature dependence of the  measured mode energies can be quantitatively described within a  multi-phonon Fr\"{o}hlich model for generic electron-phonon interactions neglecting long-range Coulomb interactions.
	\end{abstract}

	
	\maketitle
	
	\section{Introduction}
	
	The charge-density-wave (CDW) state 
	in quasi-one-dimensional conductors such as 
	K$_{0.3}$MoO$_3$ (``blue bronze'')  has received a lot of attention in the 
	1970s  and 1980s both theoretically \cite{Lee73,Lee74,Rice75,Rice76,Rice78} and           experimentally \cite{Gruener88,Gruener94,Pouget91,Hennion92,Ravy04}. 
	At a critical temperature $T_c$, CDW-materials exhibit
	a phase transition from a metallic to a semiconducting state with a gap $\Delta$ for
	electronic excitations with momenta close to the Fermi momentum $k_F$.
	The CDW state can be associated with a Peierls distortion, where phonons with
	momentum close to $q=2 k_F$ condense and, thus, generate an additional periodic potential for the electrons. 	The CDW gap $\Delta$
	is proportional to the macroscopic phonon displacement with momentum  $q = 2 k_F$.
	In this work, we consider only 
	incommensurate charge-density-waves  where
	$2 k_F$ lies (up to a reciprocal lattice vector)
	inside the first Brillouin zone. The CDW order parameter 
	is then complex so that its fluctuations can be decomposed into phase and amplitude fluctuations.
	The amplitude mode is the analog of the Higgs mode in high-energy physics, while the   
	phase mode (phason) is the Goldstone mode associated with the 
	spontaneous breaking of the $U(1)$ symmetry of the complex CDW order parameter.
	The phase mode is therefore expected to be gapless.

	Recent progress in time resolved optical and THz spectroscopy \cite{Schaefer13,Kim22,Nguyen22,Schaefer10,Schaefer14,Thomson17,Thomson23}
	has triggered renewed interest in the dynamics of collective modes in CDW systems; note in particular our companion paper \cite{Thomson23}. However, a  complete theoretical understanding of their dynamics is still not available. In particular, microscopic calculations of the damping of
	phase and amplitude modes in the CDW state as a function of temperature cannot be found 
	in the literature. In this work, we  address another relevant question which has not been completely settled: what is the effect of long-range Coulomb interactions on the spectrum of collective modes in an incommensurate CDW? Old publications addressing this problem are partially contradictory \cite{Lee78,Wong87,Virosztek93}. In particular,  Virosztek and Maki \cite{Virosztek93} found 
	that the amplitude modes are not affected by the Coulomb interaction, while the phase mode splits at finite temperature into an optical and an acoustic branch; at zero temperature  ($T=0$K) only the optical branch survives so that  Coulomb interactions destroy gapless phase fluctuations.
	In this work, we  critically re-examine this result and show
	that it is essentially modified when the Coulomb interaction between the	positively charged ions (which was neglected in Ref.~[\onlinecite{Virosztek93}])
	is taken into account. We find that in the CDW state the lowest phase mode remains gapless
	even in the presence of 
	long-range Coulomb interactions. 
	Our approach shows that the screening of the Coulomb interaction by
	acoustic phonons 
	(which is the  Goldstone mode implied by the broken translational invariance 
	in a crystal) is essential to protect the gapless nature of the lowest-frequency phase mode. 
	
	This paper is organized as follows. In Sec.~\ref{sec:models} we introduce two model Hamiltonians describing  incommensurate charge-density-waves in quasi-one-dimensional conductors: the multi-phonon Fr\"{o}hlich Hamiltonian for generic electron-phonon interactions and its extension including long-range Coulomb interactions. In Sec.~\ref{sec:MF} we recapitulate the mean-field theory for the CDW state. In the following Sec.~IV we compute the collective modes in the CDW state without long-range Coulomb interactions using an effective action approach for the generic Fr\"{o}hlich Hamiltonian within the Gaussian approximation (which is equivalent to the random-phase approximation). In Sec.~V we  derive the dispersions of collective modes including long-range Coulomb interactions and show that the lowest-frequency phase mode remains gapless. In Sec.~VI we compare our predictions for the temperature dependence  of the collective modes with experimental results for K$_{0.3}$MoO$_3$  \cite{Schaefer10,Schaefer14,Thomson17,Thomson23}. The concluding Sec.~VII gives a summary of our main results and an outlook. Additional technical details of the calculations presented in this work are given in two appendices.
	
	\section{Models}
	\label{sec:models}
	
	In this section, we introduce two model Hamiltonians  for  the theoretical description of the
	CDW state in quasi-one-dimensional conductors such as K$_{0.3}$MoO$_3$.	We also point out some subtleties related to phonon renormalization and the screening	of the Coulomb interactions which play an important role in the calculation of the collective modes in the CDW state.

	\subsection{Multi-phonon Fr\"{o}hlich model for generic electron-phonon interactions}
	
	The established  minimal model describing the CDW instability in low-dimensional conductors	is the Fr\"{o}hlich Hamiltonian for generic electron-phonon interactions \cite{Froehlich52}
	\begin{eqnarray}
		{\cal{H}}_{\rm F} & = & \sum_{\bd{k} \sigma} \epsilon_{\bd{k}} 
		c^{\dagger}_{\bd{k} \sigma} c_{\bd{k} \sigma}
		+ \sum_{\bd{q} \lambda } \omega_{\bd{q} \lambda} b^{\dagger}_{\bd{q} \lambda} 
		b_{\bd{q} \lambda }
		\nonumber
		\\
		& + & \frac{1}{\sqrt{\cal{V}}} {\sum_{\bd{k} \sigma \bd{q}  \lambda}}
		\gamma_{ \bd{q} \lambda }
		c^{\dagger}_{\bd{k} + \bd{q} \sigma} c_{\bd{k} \sigma} X_{\bd{q} \lambda},
		\label{eq:Froehlich}
	\end{eqnarray}
	where $c_{\bd{k} \sigma}$ annihilates  a spin-$\sigma$  electron with momentum 
	$\bd{k}$
	and energy $\epsilon_{\bd{k}}$, while
	$b_{\bd{q} \lambda}$ annihilates a phonon of type $\lambda$, with momentum $\bd{q}$ and energy $\omega_{\bd{q} \lambda}$. Note that the sum over $\lambda$ runs over all types of longitudinal and transverse phonons. Assuming that the Fermi surface can be approximated by two parallel flat sheets (see for example the experiments \cite{Veuillen87}) at $ \pm \bd{k}_F = \pm k_F \bd{e}_x$, where $\bd{e}_x$ is a unit vector 
	along the direction of the chains of atoms or molecules forming the  quasi-one-dimensional material,
	the dispersion of the low-energy fermionic excitations can be approximated by
	$\epsilon_{\bd{k}_F + \bd{k} } \approx \epsilon_{\bd{k}_F} + v_F k_x$, where
	$v_F$ is the Fermi velocity. In the second line of Eq.~(\ref{eq:Froehlich})
	the volume of the system is denoted by ${\cal{V}}$ and the phonon displacements are represented by the operator
	\begin{equation}
		X_{\bd{q} \lambda} = \frac{ b_{ \bd{q} \lambda} + b^{\dagger}_{-\bd{q} \lambda}}{ \sqrt{ 2
				\omega_{\bd{q} \lambda  }}}.
	\end{equation}
	We assume that the electron-phonon couplings $\gamma_{\bd{q} \lambda}$ are only finite for momenta
	$ \bd{q}$ close to $ \pm 2 \bd{k}_F$ 
	so that, to the leading order,  long-wavelength phonons with $ | \bd{q} | \ll k_F$
	are not renormalized by the interaction. As pointed out a long time ago in 
	Refs.~[\onlinecite{Brovman67,Gelikman75}] and recently emphasized in
	Ref.~[\onlinecite{Yuzbashyan22}], a finite  small-momentum part of the vertex
	$\gamma_{\bd{q} \lambda}$ in the generic Fr\"{o}hlich Hamiltonian (\ref{eq:Froehlich}) would 
	lead to a lattice instability \cite{Froehlich52} which has recently been discussed by
	several authors \cite{Kumar08,Murakami14,Ohgoe17,Esterlis18,Esterlis19,Chubukov20,Wang20,Sadovskii21,Cichutek22}. In this work we eliminate  
	possible instabilities competing with the CDW 
	by assuming that $\gamma_{\bd{q} \lambda}$ is  finite only for the momenta $\bd{q}$
	close to $\pm 2 {\bd{k}}_F$.
	
	\subsection{Fr\"{o}hlich-Coulomb model}

	To take the long-range Coulomb interaction and electron-phonon scattering
	with small momentum transfers into account, we 
	supplement the Fr\"{o}hlich Hamiltonian for generic electron-phonon interactions (\ref{eq:Froehlich})  by the quantized interaction energy 
	between all charge fluctuations \cite{Fetter71},
	\begin{eqnarray}
		{\cal{H}}_{\rm FC}  & = &  {\cal{H}}_F
		+ \frac{1}{ 2 {\cal{V}}} \sum_{\bd{q} \neq 0 } f_{\bd{q}} \rho_{ - \bd{q}} \rho_{\bd{q}},
		\label{eq:FroehlichC}
	\end{eqnarray}
	where $f_{\bd{q}}$ is the Fourier transform of the Coulomb interaction. 	For a quasi-one-dimensional system of coupled chains with transverse lattice spacing $a_{\bot}$ this gives \cite{Kopietz95,Kopietz97}
	\begin{equation}
		f_{\bd{q}} = e^2 a_{\bot}^2 \int_{ - \infty}^{\infty} d r_x \sum_{\bd{r}_{\bot}} \frac{ e^{ - i \bd{q} \cdot
				{\bd{r}}} }{ \sqrt{ r_x^2 + \bd{r}_{\bot}^2 }},
	\end{equation}
	where the $\bd{r}_{\bot}$ sum is over a two-dimensional lattice of chains and the $\bd{r}_{\bot} =0$ terms should be properly regularized.
	In the long-wavelength limit $ | \bd{q} | a_{\bot} \ll 1$ this reduces to the usual $f_{\bd{q}} = 4 \pi e^2 / \bd{q}^2$. 
	The density operator in Eq.~(\ref{eq:FroehlichC}),
	\begin{equation}
		\rho_{\bd{q}} = \rho^e_{\bd{q}} - \rho^i_{\bd{q}},
	\end{equation} 
	represents the Fourier components of the 
	total charge density consisting of the sum of the electronic density
	\begin{equation}
		\rho^e_{\bd{q}} = \sum_{ \bd{k} \sigma} c^{\dagger}_{ \bd{k} \sigma } c_{ \bd{k} + \bd{q} \sigma},
	\end{equation}
	and the ionic density
	\begin{equation}
		\rho^i_{\bd{q}} = - \sqrt{\cal{V}} \sum_{\lambda} \alpha_{\bd{q} \lambda} X_{ \bd{q} \lambda }.
	\end{equation}
	While the sum over $\lambda$ runs over all types of phonons we set $\alpha_{\bd{q} \lambda} = 0$ when $\lambda$ refers to transverse 	phonons; in contrast, the electron-phonon coupling $\gamma_{ \pm 2 \bd{k}_F  \lambda}$	in the Fr\"{o}hlich Hamiltonian (\ref{eq:Froehlich}) is non-zero for both 	longitudinal and transverse phonons. The coupling  between longitudinal phonons and the fluctuations of the ionic density for small momenta is of the form \cite{Fetter71,Mahan10}
	\begin{equation}
		\alpha_{\bd{q} \lambda} 
		= \alpha_{\lambda} | \bd{q} |,
		\label{eq:alphadef}
	\end{equation}
	where the momentum-independent constants $\alpha_{\lambda}$ depend on the phonon type.
	In particular, a  lattice with  a single  atom per unit cell supports  
	only one longitudinal acoustic phonon, so that in this case, we may omit the flavor label $\lambda$. The corresponding coupling $\alpha_{\bd{q} }$   in this case can be
	written as \cite{Fetter71}
	\begin{equation}
		\alpha_{\bd{q}} 
		=  z \sqrt{ \frac{n_i}{M} }  | \bd{q} |,
		\label{eq:alphadef1}
	\end{equation}
	where $z$ is the valence of the ions, $n_i$ is their density, and $M$ is their mass.
	Let us emphasize that the Coulomb interaction  
	in Eq.~(\ref{eq:FroehlichC}) has three contributions,
	\begin{eqnarray}
		& & \frac{1}{ 2 {\cal{V}}} \sum_{\bd{q} \neq 0 } f_{\bd{q}} \rho_{ - \bd{q}} \rho_{\bd{q}}
		=  \frac{1}{ 2 {\cal{V}}} \sum_{\bd{q} \neq 0 } f_{\bd{q}}  
		\rho^e_{ - \bd{q}} \rho^e_{\bd{q}}  
		\nonumber
		\\
		&    & 
		+ \frac{1}{ 2 {\cal{V}}} \sum_{\bd{q} \neq 0 } f_{\bd{q}}   \rho^i_{ - \bd{q}} \rho^i_{\bd{q}}  
		- \frac{1}{\cal{V}} \sum_{\bd{q} \neq 0 } 
		\rho^e_{ - \bd{q}} \rho^i_{\bd{q}} ,
		\label{eq:HCoulomb}
	\end{eqnarray}
	where the last term represents the electron-ion interaction with small momentum transfers,
	\begin{eqnarray}
		- \frac{1}{\cal{V}} \sum_{\bd{q} \neq 0 } 
		\rho^e_{ - \bd{q}} \rho^i_{\bd{q}} 
		& = & 
		\frac{1}{\sqrt{\cal{V}}} \sum_{\bd{k} \sigma \bd{q} \neq 0 , \lambda}
		f_{\bd{q}} \alpha_{\bd{q} \lambda} c^{\dagger}_{ \bd{k} + \bd{q} \sigma} c_{\bd{k} \sigma}
		X_{\bd{q} \lambda}.
		\hspace{9mm}
	\end{eqnarray}
	For acoustic phonons this is
	the usual deformation-potential coupling to electrons
	with unscreened electron-ion potential \cite{Fetter71,Mahan10}.
	For optical phonons the coupling 
	$f_{\bd{q}} \alpha_{\bd{q}} \propto 1/| \bd{q} |$ describes the polar coupling of electrons to longitudinal optical phonons \cite{Mahan10}.
	
	The ion-ion interaction on the right-hand side of  Eq.~(\ref{eq:HCoulomb}) 
	has been omitted in previous investigations of the effect of Coulomb interactions on the collective modes in CDW systems  \cite{Wong87,Virosztek93}. 
	We show here that
	for the calculation of the dispersions of the collective modes of CDW systems in the presence of Coulomb interactions, it is crucial to retain also the ion-ion interaction term
	in Eq.~(\ref{eq:HCoulomb}). 
	To give an  intuitive  argument for the importance 
	of the ion-ion interaction one can consider, for simplicity, a  
	lattice supporting only a single longitudinal acoustic
	phonon. Then the contribution from the ion-ion interaction to Eq.~(\ref{eq:HCoulomb}) can be written as
	\begin{equation}
		\frac{1}{ 2 {\cal{V}}} \sum_{\bd{q} \neq 0 } f_{\bd{q}}   \rho^i_{ - \bd{q}} \rho^i_{\bd{q}}  =
		\frac{1}{2} \sum_{\bd{q} \neq 0 } \Omega_i^2 X_{ - \bd{q}} X_{\bd{q}},
		\label{eq:cbii}
	\end{equation}
	where we have used the fact that the combination
	\begin{equation}
		f_{\bd{q}} \alpha_{\bd{q}}^2 = \Omega_i^2 = 4 \pi ( z e )^2 n_i /M
		\label{eq:ionicplasma}
	\end{equation}
	can be identified with the square of the ionic plasma frequency.
	At the first glance, Eq.~(\ref{eq:cbii}) seems to suggest that long-range Coulomb
	interactions push the frequency of acoustic phonons
	up to the ionic plasma frequency. In a metal this is of course incorrect, because
	the ionic charge is screened by the electrons so that acoustic phonons have the squared dispersion $ \Omega_i^2 / ( 1 + \kappa^2 / q^2 )$, where $\kappa$ is the Thomas-Fermi screening wavevector. For $ q \ll \kappa$ we thus recover the  linear dispersion of acoustic phonons with velocity $ \Omega_i / \kappa$,  which is the well-known  Bohm-Staver relation \cite{Bohm51,Ashcroft76}.
	From the above considerations, it should be clear that by simply dropping the ionic contribution to the Coulomb interaction in Eq.~(\ref{eq:HCoulomb}) one violates the balance between electronic and ionic charge fluctuations and therefore cannot properly describe screening effects.
	To calculate the effect of Coulomb interactions on the collective modes in CDW systems it is,
	therefore, crucial to retain also the ionic contribution to the Coulomb Hamiltonian (\ref{eq:HCoulomb}). 
	In Sec.~\ref{sec:Coulomb} we will calculate the  energies of amplitude and phase modes including the effect of Coulomb interactions. There we find that the ionic part in Eq.~(\ref{eq:HCoulomb}) leads to a new contribution to the collective modes energies which can have the same order of magnitude as the result obtained in Ref.~[\onlinecite{Virosztek93}] 	where the purely ionic part in Eq.~\eqref{eq:HCoulomb} was neglected. 	It turns out that this ionic part is crucial to describe the complete screening of electronic charge fluctuations by the ions, which in turn protects the gapless nature of the Goldstone mode associated with the broken $U(1)$ symmetry in the CDW state. 
	
	\begin{figure}[tb]
		\centering
		\includegraphics[width=0.45\textwidth]{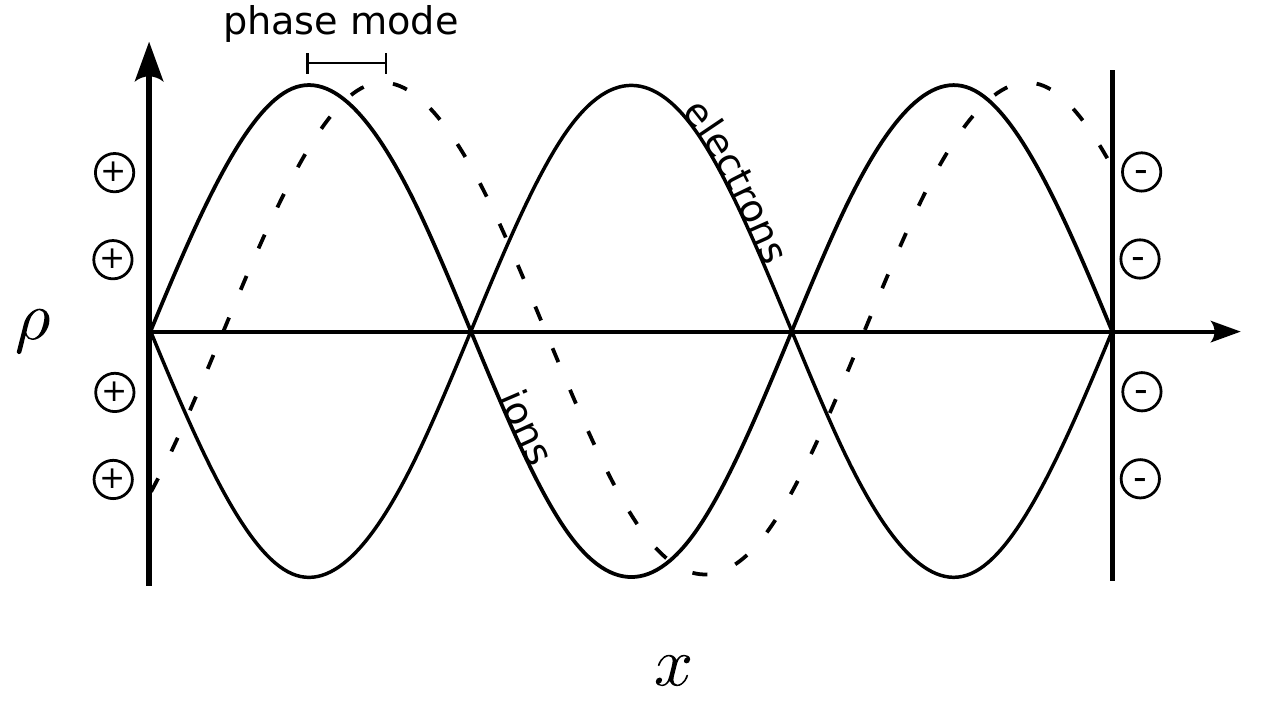}
		\caption{Formation of the collective phase mode without the Coulomb interaction between the ions. The full lines marked ``electrons'' and ``ions'' are the  profiles of the electronic and the ionic densities in the CDW state respectively. The dashed line is the electronic density profile for the phase mode, in which its phase acquires a shift with respect to the lowest energy state. The plus and minus charges mark formation of a macroscopic dipole in the system when the phase mode is excited, see also chapter 6 of Ref. \cite{Grunerbook94}.}
		\label{fig:fig_phason_ii_int}
	\end{figure}

	This effect of the ionic part on the formation of Goldstone (phase) mode can be understood intuitively on the physical level as follows. The CDW state is formed due to the Peiels instability, {\it i.e.},  the electrons and ions form a wave state in their respective densities at a given momentum with the relative phase shift of $\pi$ so that the net charge remains zero everywhere throughout the system, as illustrated by the two solid lines in Fig.\ \ref{fig:fig_phason_ii_int}. An excitation of the phase type above this state is a deviation in the  phase of the electronic density profile, see the dashed line in Fig.~\ref{fig:fig_phason_ii_int}, that produces a  macroscopic dipole moment, which costs a finite charging energy since the Coulomb force is long-range and results in a finite gap in the spectrum of the phase excitations. Turning on the Coulomb interaction between the ions allows the screening processes for electrons by the ions. Such a process is also invoked in the intuitive derivation of the Bohm-Staver relation \cite{Bohm51,Ashcroft76}. Such processes also allow  the ionic density profile to follow the electron one in Fig.\ \ref{fig:fig_phason_ii_int} resulting in absence of any macroscopic dipole and eliminating, therefore, the gap in the spectrum fo the phase mode. In a different context, the importance of the ion-ion contribution has been noted previously in Refs.~[\onlinecite{Kopietz96a,Kopietz96b}].
	
	\section{Mean-field theory for the CDW}
	\label{sec:MF}
	
	To fix our notation and set the stage for the calculation of the collective modes, let us briefly 
	recall the usual mean-field theory for the CDW state in a quasi-one-dimensional 
	metal \cite{Gruener94} within a functional integral approach. 
	Starting point is the Euclidean action of the Fr\"{o}hlich 
	model for generic electron-phonon interactions defined in Eq.~(\ref{eq:Froehlich}),
	\begin{eqnarray}
		S_F [ \bar{c} , c , X ] & = & - \int_K \sum_{\sigma} G_0^{-1} ( K ) \bar{c}_{ K \sigma}  c_{K \sigma}
		\nonumber
		\\
		& + &  \frac{1}{2} \int_Q \sum_{\lambda} D_{0, \lambda} ^{-1} ( Q) X_{- Q } X_Q 
		\nonumber
		\\
		& + & \int_{K} \int_Q \sum_{\sigma \lambda} \gamma_{\bd{q} \lambda} 
		\bar{c}_{K + Q \sigma} c_{K \sigma} X_{Q \lambda},
		\label{eq:SF}
	\end{eqnarray}
	where the inverse propagators of the electrons and phonons are
	\begin{subequations}
		\begin{eqnarray}
			G_0^{-1} ( K) & = & i \omega - \epsilon_{\bd{k}} + \mu,
			\\
			D_{0 , \lambda}^{-1} ( Q ) & = & \nu^2 + \omega_{\bd{q} \lambda}^2 .		\end{eqnarray}
	\end{subequations}
	Here the collective label $ K = ( i \omega , \bd{k} )$ represents fermionic Matsubara frequency $i \omega $ and momentum $\bd{k}$, the label
	$Q = ( i \nu , \bd{q} )$ represents bosonic Matsubara frequency $ i \nu$ and momentum $\bd{q}$, and the integration
	symbols are defined by
	$\int_K = ( \beta {\cal{V}} )^{-1} \sum_{\bd{k}} \sum_{\omega}$ and
	$\int_Q = ( \beta {\cal{V}} )^{-1} \sum_{\bd{q}} \sum_{\nu}$, where $\beta = 1/T$ is the inverse temperature.
	
\begin{figure}[tb]
		\centering
		\includegraphics[width=0.45\textwidth]{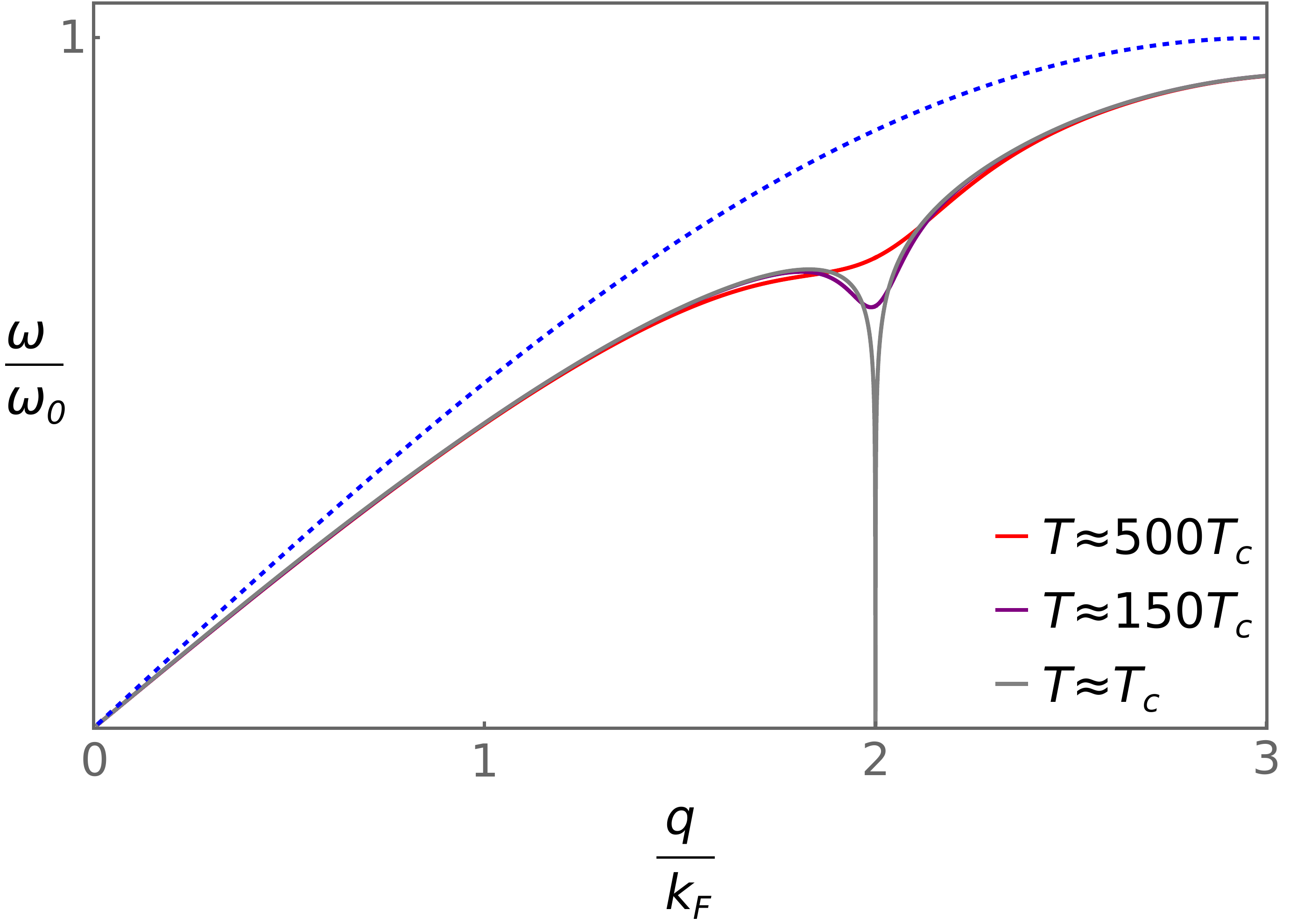}
		\caption{Schematic visualization of the softening  of phonons with  wavevector $q =2k_F$
			as the temperature $T$ approaches the critical temperature $T_c$ 
			of the CDW instability.
			The dashed-blue line denotes a generic 
			one-dimensional phonon 
			dispersion  $\omega_q=2\omega_0\sin(qa/2)$, while the solid lines denote the renormalized phonon dispersion $\tilde{\omega}_q=\omega_q\sqrt{1-\lambda_0\Pi_0(q,0)/\Pi_0(0,0)}$ for different temperatures. Here $\lambda_0$ represents a (constant) dimensionless electron-phonon coupling and $\Pi_0(q,0)$ is the static 
			Lindhard function in one dimension  in the normal 
			phase \cite{Mihaila11}. Because $\Pi_0 ( 2 k_F, 0 )$  
			diverges logarithmically for $T \rightarrow 0$, 
			there is a finite critical temperature $T_c$ where
			$\tilde{\omega}_{2k_F}$ vanishes. At lower temperatures
			a CDW with periodicity 
			${Q}_{*}=2 {k}_F$ emerges.}
		\label{fig:fig0}
	\end{figure}

	To investigate the CDW instability within the 
	mean-field approximation, we replace the phonon displacement field
	by its expectation value describing  a CDW with 
	ordering wave-vector $ \bd{Q}_{\ast} = 2 \bd{k}_F$,
	\begin{equation}
		X_{Q \lambda} \rightarrow  X^0_{Q \lambda} = \beta  {\cal{V}}  \delta_{ \nu , 0}   [   
		\delta_{\bd{q} ,  \bd{Q}_{*} }  {X}_{ \bd{Q}_{*} \lambda}^0 
		+   \delta_{\bd{q} , -  \bd{Q}_{*} }  {X}^0_{- \bd{Q}_{*} \lambda} ],
	\end{equation}
	where $\bd{Q}_{*} = 2 \bd{k}_F$ is the ordering momentum of the CDW. A sketch of  the softening of the phonon dispersion at wave-vector $q = 2 k_F$ when the temperature approaches the critical CDW temperature $T_c$ is shown in Fig.~\ref{fig:fig0}.
	With the notation
	
	\begin{equation}
		\Delta = \sum_{\lambda} \gamma_{ \bd{Q}_{*} \lambda} X^0_{\bd{Q}_{*} \lambda},
		\; \; \; 
		\Delta^{\ast} = \sum_{\lambda} \gamma_{ -\bd{Q}_{*} \lambda} X^0_{- \bd{Q}_{*} \lambda}
	\end{equation}
	the mean-field action can be written as
	\begin{eqnarray}
		S_F [ \bar{c} , c , X^0 ] & = &  \int_K \sum_{\sigma}
		\Bigl[
		(\epsilon_{\bd{k}}  - \mu  -  i \omega )  \bar{c}_{ K \sigma}  c_{K \sigma}
		\nonumber
		\\
		& + &   \Delta \bar{c}_{K +  \bd{Q}_{*} \sigma} 
		c_{K \sigma} +   \Delta^{\ast} \bar{c}_{K -  \bd{Q}_{*} \sigma} 
		c_{K \sigma}  \Bigr] 
		\nonumber
		\\
		& + &
		\beta {\cal{V}} \sum_{\lambda}  \omega_{ \bd{Q}_{*} \lambda}^2 
		X^0_{ - \bd{Q}_{*} \lambda} X^0_{ \bd{Q}_{*} \lambda}.
		\label{eq:SF0}
	\end{eqnarray}
	This is the action 
	of non-interacting electrons moving in an additional  
	periodic potential 
	$
	U ( \bd{r} ) =  \Delta  e^{i \bd{Q}_{*} \cdot \bd{r}}
	+ \Delta^{\ast} e^{- i \bd{Q}_{*} \cdot \bd{r}}$.
	The corresponding eigenstates are Bloch states
	and the spectrum consists of infinitely many energy bands $E_{\bd{k} n }$,
	where $ \bd{k}$ is restricted to the first Brillouin zone and $n$ enumerates the bands.
	With  $\bd{Q}_{*} = 2 k_F  \bd{e}_x$ the first Brillouin zone is  $ | k_x | \leq k_F$.
	Since we are only interested in the low-energy states we retain only the lowest two energy bands by restricting the momentum sum to the regime $  |  k_x | \leq  2 k_F$
	consisting of the first two Brillouin zones.
	After shifting the momentum labels in the anomalous terms 
	the mean-field action reduces to
	\begin{widetext}
		\begin{eqnarray}
			S_F [ \bar{c} , c , X^0] & = &  \int_{K} \sum_{ \sigma}
			\left(  \bar{c}_{ K + \frac{\bd{Q}_{*}}{2} \sigma } ,  \bar{c}_{ K - \frac{\bd{Q}_{*}}{2} \sigma } 
			\right)
			\left(  \begin{array}{cc} \epsilon_{\bd{k} + \frac{\bd{Q}_{*}}{2} }  - \mu  - i \omega  &  \Delta \\
				\Delta^{\ast} &   \epsilon_{\bd{k} - \frac{\bd{Q}_{*}}{2} }   - \mu  - i \omega \end{array} \right)
			\left( \begin{array}{c}  c_{ K + \frac{\bd{Q}_{*}}{2} \sigma } \\
				c_{K - \frac{\bd{Q}_{*}}{2} \sigma } \end{array} \right)
			+ \beta {\cal{V}} \sum_{\lambda} \omega_{\bd{Q}_{*} \lambda}^2 X^0_{- \bd{Q}_{*} \lambda} 
			X^0_{\bd{Q}_{*} \lambda},
			\nonumber
			\\
			& &
			\label{eq:SFMF}
		\end{eqnarray}
	\end{widetext}
	where  the momentum integration is restricted to $ | k_x | <  k_F$ and we have introduced the notation
	$ K \pm \bd{Q}_{*} /2 = ( i \omega , \bd{k} \pm \bd{Q}_{*} / {2}) $.
	The quadratic form in Eq.~(\ref{eq:SFMF})
	can be diagonalized with a canonical  transformation to a new set of fermionic fields $d^+_{K  \sigma }$ and
	$d^-_{K  \sigma  }$,
	\begin{equation}
		\left( \begin{array}{c}  c_{ K + \frac{\bd{Q}_{*}}{2} \sigma } \\
			c_{K - \frac{\bd{Q}_{*}}{2} \sigma } \end{array} \right) =
		\left( \begin{array}{cc} u_{\bd{k}} & - v_{\bd{k}} \\
			v_{\bd{k}}^{\ast} & u^{\ast}_{\bd{k}} \end{array} \right)
		\left( \begin{array}{c}  d^+_{ K   \sigma  } \\
			d^-_{ K  \sigma  }  \end{array} \right),
	\end{equation}
	where
	\begin{eqnarray}
		u_{\bd{k}} & = & \frac{ \Delta }{ | \Delta | } \sqrt{ \frac{ E_{\bd{k}} + \xi_{\bd{k}} }{ 2 E_{\bd{k}} }}, \; \; \; 
		v_{\bd{k}}  =   \sqrt{ \frac{ E_{\bd{k}} - \xi_{\bd{k}} }{ 2 E_{\bd{k}} }},
		\\
		E_{\bd{k}} & = & \sqrt{ \xi_{\bd{k}}^2 + | \Delta |^2 } ,
		\; \; \; 
		\xi_{\bd{k}}  =  \frac{ \epsilon_{\bd{k} + \frac{\bd{Q}_{*}}{2} } -    \epsilon_{\bd{k} - \frac{\bd{Q}_{*}}{2} }}{2}.
	\end{eqnarray}
	In terms of the new fermion fields the
	mean-field action (\ref{eq:SFMF}) can be written as
	\begin{eqnarray}
		S_F [ \bar{d} , d, X^0 ]  & = &  \int_K \sum_{ \sigma} \sum_{ \alpha = \pm} 
		( E_{\bd{k}}^{ \alpha} - \mu - i \omega ) \bar{d}^{\alpha}_{K  \sigma  } d^{\alpha}_{K  \sigma }
		\nonumber
		\\
		& + &    \beta {\cal{V}} \sum_{\lambda}
		\omega_{\bd{Q}_{*} \lambda}^2 X^0_{- \bd{Q}_{*} \lambda} X^0_{\bd{Q}_{*} \lambda} ,	\end{eqnarray}
	where
	\begin{equation}
		E_{\bd{k}}^{ \pm} =   \pm E_{\bd{k}} 
		+   \frac{ \epsilon_{\bd{k} + \frac{\bd{Q}_{*}}{2} } +    \epsilon_{\bd{k} - \frac{\bd{Q}_{*}}{2} }}{2}.
	\end{equation}
	Note that the original energy dispersion $\epsilon_{\bd{k}}$  is split into a 
	conduction band  $E_{\bd{k}}^+$ and a valence band $E_{\bd{k} }^-$ separated by a gap $2 | \Delta |$.
	The corresponding mean-field
	grand canonical potential is 
	\begin{eqnarray}
		\Omega_{\rm MF} & = & 
		- \frac{s}{ \beta} \sum_{\bd{k},  \alpha = \pm} 
		\ln [ 1 + e^{ - \beta ( E_{\bd{k}}^{ \alpha} - \mu ) } ]
		\nonumber
		\\
		& +  &   {\cal{V}} \sum_{\lambda} \omega_{\bd{Q}_{*} \lambda}^2  
		X^0_{- \bd{Q}_{*} \lambda} X^0_{\bd{Q}_{*} \lambda} ,
	\end{eqnarray}
	where  
	\begin{equation}
		s =  2S+1 = 2
	\end{equation}
	is the spin degeneracy.
	Minimizing $\Omega_{\rm MF}$  
	with respect  $X^0_{ - \bd{Q}_{*}}$ we obtain the self-consistency  conditions
	\begin{equation}
		X^0_{\bd{Q}_{*} \lambda}  =  - \frac{ \gamma^{\ast}_{\bd{Q}_{*} \lambda}}{
			\omega^2_{\bd{Q}_{*} \lambda} }\frac{ s}{{\cal{V}}} \sum_{\bd{k} \alpha \lambda^{\prime}} 
		\frac{ \alpha}{ e^{\beta   ( E^{\alpha}_{\bd{k}} - \mu )  }+1}
		\frac{ \gamma_{\bd{Q}_{*} \lambda^{\prime}} X^0_{\bd{Q}_{*} \lambda^{\prime}} }{2 E_{\bd{k}}}.
		\label{eq:selfconXs}
	\end{equation}
	Keeping in mind that we consider electronically quasi-one-dimensional systems we may expand the electronic energy dispersion to linear order around the Fermi momentum, $\epsilon_{ \bd{k} \pm \frac{\bd{Q}_{*}}{2} } =	\epsilon_{ \bd{k} \pm \bd{k}_F } \approx   \epsilon_{\bd{k}_F} \pm  v_F k_x \equiv \epsilon_F\pm  v_F k_x $, so that
	\begin{equation} 
		E^{\pm}_{\bd{k}} - \mu  = \pm \sqrt{ ( v_F k_x )^2 + | \Delta |^2 } 
		\equiv \pm E_{\bd{k}}.
		\label{eq:Ekdef}
	\end{equation}
	The linearization is valid when the CDW gap $\left|\Delta\right|$ is much smaller than the Fermi energy $\epsilon_F$, so that the behavior of the system is dominated by low-energy excitations around the Fermi level.

\begin{figure}[tb]
		\begin{center}
			\centering
			\includegraphics[width=0.45\textwidth]{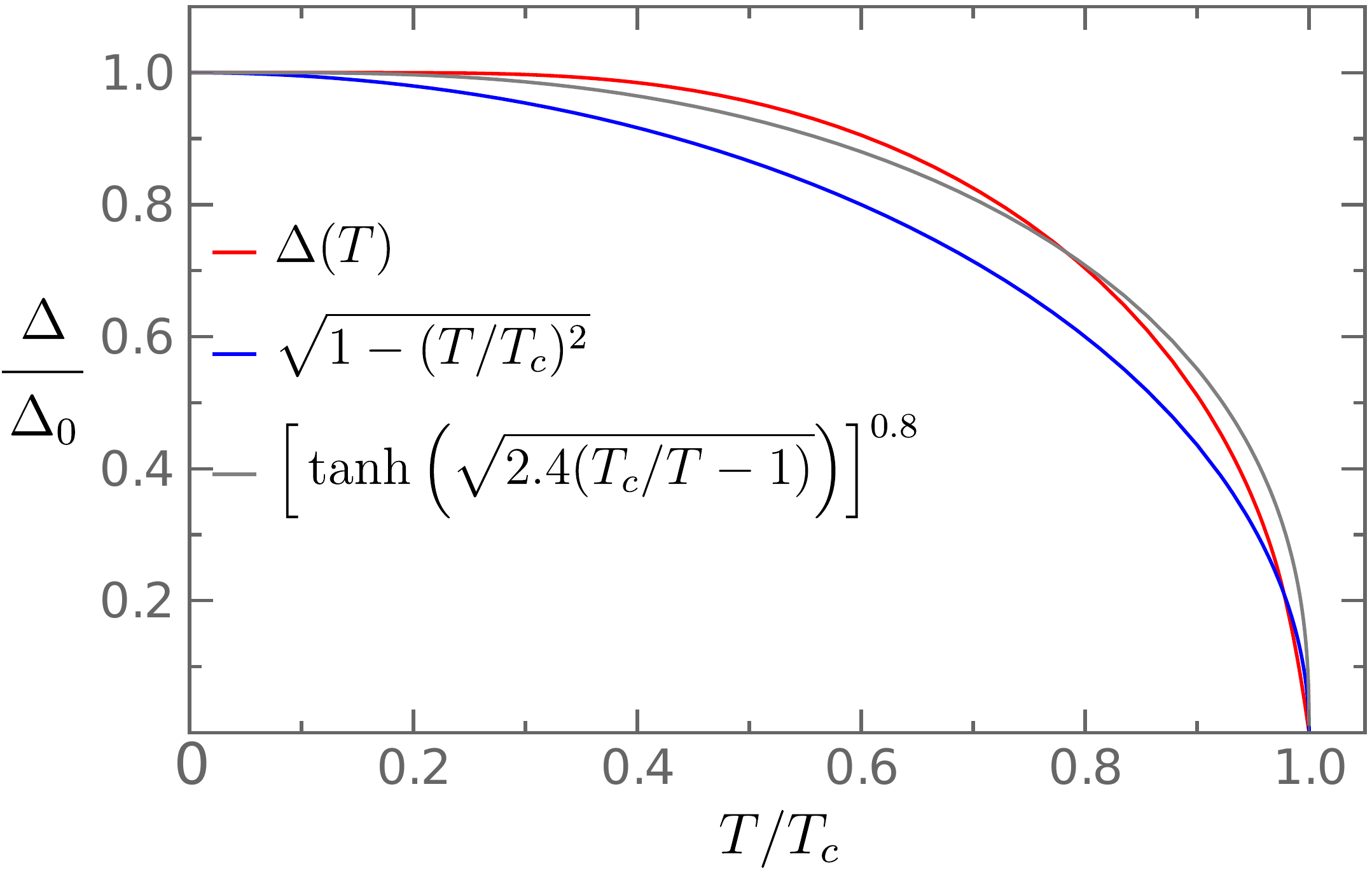}
		\end{center}
		\caption{Different expressions for the temperature dependent CDW gap $\Delta(T)$ in units of 
			$\Delta_0 = \Delta(T=0)$. The red curve is the prediction of the mean-field 
			gap equation~\eqref{eq:selfcon} for $g_0\approx0.82$, which is the sum of the relevant electron-phonon couplings for blue bronze (see Table~\ref{fig:werte} in 
			Appendix~B).
			The blue curve is the square-root behavior obtained within Ginzburg-Landau theory.
			Lastly, the  gray curve is an empirical formula fitting the measurements of Ref.~[\onlinecite{Girault89}] 
			on blue bronze.}
		\label{fig:Gap}
	\end{figure}

	Defining $\Delta_{\lambda} = \gamma_{\bd{Q}_{*} \lambda} X^0_{\bd{Q}_{*} \lambda }$ the self-consistency equation (\ref{eq:selfconXs}) can then be written as
	\begin{equation}
		\Delta_{\lambda} = d_{\lambda}  \sum_{\lambda^{\prime}} \Delta_{\lambda^{\prime}},
		\label{eq:eigenv} 
	\end{equation}
	where we have introduced the dimensionless couplings
	\begin{eqnarray}
		d_{\lambda} &  = &  -  \frac{ | \gamma_{\bd{Q}_{*} \lambda} |^2 }{
			\omega_{\bd{Q}_{*} \lambda}^2 } 
		\frac{ s }{\cal{V}} \sum_{\bd{k} \alpha}
		\frac{ \alpha}{ e^{\beta ( E_{\bd{k}}^{\alpha} - \mu )}+1} 
		\frac{ 1}{ 2 E_{\bd{k}}}
		\nonumber
		\\
		& = & \frac{ | \gamma_{\bd{Q}_{*} \lambda} |^2 }{
			\omega_{\bd{Q}_{*} \lambda}^2 } 
		\frac{ s }{\cal{V}} \sum_{\bd{k} }
		\frac{ \tanh ( \beta E_{\bd{k}} /2 ) }{ 2 E_{\bd{k}}}.
	\end{eqnarray}
	The non-zero solution of the eigenvalue equation (\ref{eq:eigenv})
	is of the form
	$\Delta_{\lambda} = \Delta d_{\lambda}$ where the parameter 
	$\Delta$ is determined by the self-consistency condition
	\begin{equation}
		1 =   \sum_{\lambda} d_{\lambda} =   \left[ \sum_{\lambda} \frac{ | \gamma_{\bd{Q}_{*} \lambda} |^2 }{
			\omega_{\bd{Q}_{*} \lambda}^2 } \right]
		\frac{ s }{\cal{V}} \sum_{\bd{k} } 
		\frac{ \tanh ( \beta E_{\bd{k}} / 2 ) }{ 2 E_{\bd{k}}}.
		\label{eq:selfcon}
	\end{equation}
	Note that the couplings $d_{\lambda}$ implicitly depend on
	$\Delta$ via $E_{\bd{k}} = \sqrt{ ( v_F k_x )^2 + | \Delta |^2}$.
	The self-consistency condition
	$\Delta = \sum_{\lambda} \Delta_{\lambda} = \Delta 
	\sum_{\lambda} d_{\lambda}$ is the multi-phonon generalization of the well known 
	mean-field self-consistency condition for the CDW order-parameter \cite{Gruener94}.
	For an electronically one-dimensional system where  $E_{\bd{k}}$ is given 
	by Eq.~(\ref{eq:Ekdef}) 
	the self-consistency equation (\ref{eq:selfcon}) reduces to
	\begin{equation}
		\frac{1}{g_0} = \int_0^{\omega_c} \frac{ dE}{ \sqrt{ E^2 + | \Delta |^2}}
		= \frac{1}{2} \ln \left( \frac{ \sqrt{ 1 + | \Delta |^2/ \omega_c^2 } + 1 }{
			\sqrt{ 1+ | \Delta |^2/ \omega_c^2 } - 1 } \right),
		\label{eq:gapequation}
	\end{equation}
	where $\omega_c$ is an ultraviolet cutoff \cite{Fetter71,Mahan10,Khomskii10} typically taken to be of the order of the bandwidth \cite{Rice75,Gruener88} and
	we have introduced the dimensionless electron-phonon coupling
	\begin{equation}
		g_0 =  \left[ \sum_{\lambda} \frac{ | \gamma_{\bd{Q}_{*} \lambda} |^2 }{
			\omega_{\bd{Q}_{*} \lambda}^2 } \right] \frac{\nu_3 }{2}.
		\label{eq:g0def}
	\end{equation} 
	Here
	\begin{equation}
		\nu_3 =  \frac{s}{\cal{V}} \sum_{\bd{k}} 2 \delta ( v_F k_x ) = \frac{s}{\pi v_F a_{\bot}^2},
		\label{eq:nu3def}
	\end{equation}
	is the (three-dimensional) density of states at the Fermi energy (including the
	spin degeneracy) of an electronically one-dimensional system with transverse lattice spacing $a_{\bot}$. 
	The solution of the self-consistency equation (\ref{eq:gapequation}) is
	\begin{equation}
		\Delta = \frac{2 \omega_c}{ e^{ 1 / g_0} - e^{ - 1 / g_0}}\label{Delta_order_parameter}.
	\end{equation}
	For small $g_0$ this can be approximated by the usual exponentially small BCS result
	$|  \Delta | \approx 2 \omega_c e^{ -1/ g_0 }$. In the opposite limit of large $g_0$ we obtain $ | \Delta | \approx \omega_c g_0$, but in this regime our  
	approximations (such as the linearization of the energy dispersion) are not valid
	so that, in the rest of this work, we assume $g_0 \lesssim 1$. 
	In Fig.~\ref{fig:Gap} we show a graph of the temperature dependence of the mean-field gap predicted by Eq.~\eqref{eq:selfcon} in comparison with two 
	additional curves: the square root behavior predicted by 
	Ginzburg-Landau theory  and an empirical formula fitting the 
	measurements of Ref.~[\onlinecite{Girault89}] on blue bronze.

	\section{Collective modes for the
		Fr\"{o}hlich model}
	\label{sec:modesF}

	To obtain the spectrum of collective modes within the Gaussian approximation,
	we write down the Euclidean action of our model and integrate over the electrons and
	the phonon momenta conjugate to the phonon displacements $X_{\bd{q} \lambda}$.
	Retaining only the two electronic bands in the vicinity of the Fermi energy
	the resulting effective phonon action is
	\begin{eqnarray}\label{eq:Seffstart}
		S_{\rm eff} [ X ] & = & \frac{1}{2} \int_{ Q} \sum_{\lambda}
		( \nu^2 + \omega^2_{\bd{q} \lambda } ) X_{ - Q \lambda } X_{ Q \lambda }
		\nonumber
		\\
		& & - s {\rm Tr} \ln [ \mathbf{G}_0^{-1} - \mathbf{\Delta } ],
	\end{eqnarray}
	where for an electronically  one-dimensional system we may linearize the electronic dispersion close to the two Fermi points so that the inverse electronic propagator in the normal state is
	\begin{eqnarray}
		[ \mathbf{G}_0^{-1} ]_{ K K^{\prime}} & = & \delta_{ K , K^{\prime}}
		\left( \begin{array}{cc} i \omega - v_F k_x & 0 \\
			0  &  i \omega + {v}_F {k}_x \end{array} \right),
		\hspace{7mm}
	\end{eqnarray}
	with the prefactor $\delta_{K , K^{\prime} } = \beta {\cal{V}}
	\delta_{ \bd{k} , \bd{k}^{\prime}} \delta_{ \omega , \omega^{\prime}}$.
	Recall that $K = ( \bd{k} , i \omega ) $ and $ Q = ( \bd{q} , i \nu )$
	are collective labels for the momenta and  Matsubara frequencies
	of the fermions and phonons, respectively.
	The electron-phonon coupling is described by the matrix
	\begin{eqnarray}
		{[} \mathbf{\Delta} {]}_{ K K^{\prime}} & = & 
		\left( \begin{array}{cc} 0  & 
			\Delta_{ K - K^{\prime}} \\
			\Delta^{\ast}_{ K^{\prime} -K} & 0 \end{array} \right),
		\nonumber
		\\
		& &
	\end{eqnarray}
	where
	\begin{equation}
		\Delta_Q = \sum_{\lambda} \gamma_{ 2 \bd{k}_F + Q \lambda} X_{ 2 \bd{k}_F + Q \lambda}.
	\end{equation}
	Separating the mean-field part of the fluctuating gap as
	$\Delta_Q = \Delta_Q^0 + V_Q$, where $\Delta^0_Q = 
	\delta_{Q,0} \Delta = 
	\beta {\cal{V}} \delta_{\nu,0}
	\delta_{\bd{q},0} \Delta$,
	we obtain, within the Gaussian approximation,
	\begin{eqnarray}
		& &  - s{\rm Tr} \ln [ \mathbf{G}_0^{-1} - \mathbf{\Delta } ] =
		- s {\rm Tr} \ln [ \mathbf{G}_{\rm 1}^{-1} - \mathbf{V } ]
		\nonumber
		\\
		&  & = \beta \Omega_{1} + s {\rm Tr} [ \mathbf{G}_{1} \mathbf{V} ]
		+ \frac{ s}{2} {\rm Tr} [ \mathbf{G}_{1} \mathbf{V}  \mathbf{G}_{1} \mathbf{V}  ]   
		+ {\cal{O}} ( V^3 ),
		\hspace{7mm} \label{eq:expansion1}
	\end{eqnarray}
	where
	\begin{eqnarray}
		\beta \Omega_1 & = &  - s {\rm Tr} \ln \mathbf{G}_1^{-1}
		=   - s \sum_{\bd{k} \alpha} \ln [ 1 + e^{ - \beta ( E_{\bd{k}}^{ \alpha} - \mu ) } ],
		\hspace{7mm}
	\end{eqnarray}
	is the electronic contribution to the mean-field grand canonical potential in units of temperature. The inverse electronic 	mean-field propagator is given by
	\begin{eqnarray}
		[ \mathbf{G}_1^{-1} ]_{ K K^{\prime}} & = & \delta_{ K , K^{\prime}}
		\left( \begin{array}{cc} i \omega - v_F k_x & \Delta \\
			\Delta^{\ast}  &  i \omega + {v}_F {k}_x \end{array} \right), \phantom{a}
		\label{eq:G1def}
	\end{eqnarray}
	and the matrix $\mathbf{V}$ contains the fluctuation 
	$V_Q = \Delta_Q - \Delta^0_Q$
	of the CDW order parameter,
	\begin{eqnarray}
		{[} \mathbf{V} {]}_{ K K^{\prime}} & = & 
		\left( \begin{array}{cc} 0  & 
			V_{ K - K^{\prime}} \\
			V^{\ast}_{ K^{\prime} -K} & 0 \end{array} \right).
		\\
		& & \nonumber
	\end{eqnarray}
	Note that this matrix has no diagonal elements because we have assumed that
	the electron-phonon vertex $\gamma_{\bd{q} \lambda }$
	in the Fr\"{o}hlich Hamiltonian for generic electron-phonon interactions is only finite for momenta close to
	$\pm 2 \bd{k}_F$.
	To explicitly separate the long-wavelength phonon displacements with momenta $ | \bd{q} | \ll k_F$
	from the  short-wavelength displacements with momenta close to $ \pm 2 \bd{k}_F$  (see Fig.~\ref{fig:fig0}), we set
	\begin{subequations}
		\label{eq:UVlow}
		\begin{eqnarray}
			U_{Q \lambda } & = & X_{ Q \lambda }, \; \; \; \mbox{$ | \bd{q} | \ll k_F $},
			\\
			V_{ Q \lambda} & = &  X_{ 2 \bd{k}_F + Q \lambda} - \delta_{ Q,0} X^{0}_{ 2 \bd{k}_F \lambda },  \; \; \; \mbox{$ | \bd{q} | \ll k_F $}.
			\hspace{7mm}
		\end{eqnarray}
	\end{subequations}
	Note that  by definition $U^{\ast}_{  Q \lambda } = U_{ - Q \lambda }$ but for an incommensurate CDW
	$V^{\ast}_{ Q \lambda } = X_{ - 2 \bd{k}_F - Q , \lambda } -
	\delta_{Q,0} X^0_{ - 2 \bd{k}_F , \lambda} \neq V_{ -Q \lambda }$.
	With this notation, we write the fluctuations of the gap as 
	\begin{eqnarray}
		V_Q & = &  \Delta_Q - \Delta_Q^0 =  \sum_{\lambda} \gamma_{ 2 \bd{k}_F + \bd{q} \lambda}  V_{ Q \lambda}.
	\end{eqnarray}
	The self-consistency equations (\ref{eq:selfconXs}) guarantee that the
	expansion of $S_{\rm eff} [  X ]$ does not have a linear term in the fluctuations, so that in Gaussian approximation the effective action Eq.~\eqref{eq:Seffstart}  reduces to 
	\begin{eqnarray}
		S_{\rm eff} [  U , V  ] & = & \beta \Omega_1 + \beta {\cal{V}} \sum_{\lambda} \omega_{ 2 \bd{k}_F  \lambda}^2 X^{0}_{ - 2 \bd{k}_F \lambda} X^0_{ 2 \bd{k}_F \lambda} 
		\nonumber
		\\
		& + &  \frac{1}{2} \int_{ Q} \sum_{\lambda}
		( \nu^2 + \omega^2_{\bd{q} \lambda } ) U_{ - Q \lambda } U_{ Q \lambda }
		\nonumber
		\\
		& + & \int_{ Q} \sum_{\lambda}
		( \nu^2 + \omega^2_{2 \bd{k}_F + \bd{q} \lambda } ) V^{\ast}_{  Q \lambda } V_{ Q \lambda } 
		\nonumber
		\\
		& + &  \frac{ s}{2} {\rm Tr} [ \mathbf{G}_{1} \mathbf{V}  \mathbf{G}_{1} \mathbf{V}  ]  .
		\label{eq:SeffuV}
	\end{eqnarray}
	The trace in the last term can be written as
	\begin{eqnarray}
		& & \frac{s}{2} {\rm Tr} [ \mathbf{G}_{1} \mathbf{V}  \mathbf{G}_{1} \mathbf{V}  ] 
		= \int_Q \Bigl[ \Pi_0^{ \bar{V} V } ( Q ) V^{\ast}_Q V_Q
		\nonumber
		\\
		& & 
		+ \frac{ \Pi_0^{VV} ( Q ) }{2} V_{-Q } V_Q
		+ \frac{ \Pi_0^{\bar{V} \bar{V}} ( Q ) }{2} V^{\ast}_{Q } V^{\ast}_{-Q}		\Bigr].
	\end{eqnarray}
	where we have introduced the following three  polarization functions,
	\begin{widetext}
		\begin{subequations}
			\label{eq:pidefs}
			\begin{eqnarray}
				\Pi_0^{\bar{V} V} ( \bd{q} , i \nu  ) & = & s \int_K
				\frac{ i \omega ( i \omega + i \nu ) -  v_F^2 k_x ( k_x +q_x ) }{
					[ ( i \omega )^2 - ( v_F k_x )^2 - | \Delta |^2 ]
					[ ( i \omega + i \nu )^2 - ( v_F (k_x + q_x))^2 - | \Delta |^2 ] },
				\\
				\Pi_0^{VV} ( \bd{q} , i \nu  ) & = & s
				\int_K
				\frac{ (\Delta^{ \ast})^2 }{
					[ ( i \omega )^2 - ( v_F k_x )^2 - | \Delta |^2 ]
					[ ( i \omega + i \nu )^2 - ( v_F (k_x + q_x))^2 - | \Delta |^2 ] },
				\\
				\Pi_0^{ \bar{V} \bar{V}} ( \bd{q} , i \nu  ) & = & s
				\int_K \frac{ \Delta^2 }{
					[ ( i \omega )^2 - ( v_F k_x )^2 - | \Delta |^2 ]
					[ ( i \omega + i \nu )^2 - ( v_F (k_x + q_x))^2 - | \Delta |^2 ] }.
			\end{eqnarray}
		\end{subequations}
	\end{widetext}
	Note that in the effective action \eqref{eq:SeffuV} the long-wavelength phonons with momenta $ | \bd{q} | \ll k_F$ decouple from 	the phonons with momenta close to $ \pm 2 \bd{k}_F$, i.e., the two types of
	phonons do not interact within this approximation. \\
	
	For simplicity, let us use the gauge freedom to choose the mean-field order parameter $\Delta$ to be real in order to simplify our calculations. It is then convenient to parametrize the complex field $V_{Q}$ in terms of two real fields
	$A_{Q}$ and $B_{Q}$ by setting
	\begin{subequations}
		\label{eq:ABdef}
		\begin{eqnarray}
			V_Q & = & \frac{1}{\sqrt{2}} [ A_Q + i B_Q ],
			\\
			V^{\ast}_Q & = & \frac{1}{\sqrt{2}} [ A_{-Q} - i B_{-Q} ].
		\end{eqnarray}
	\end{subequations}
	With this notation
	\begin{eqnarray}
		& & \frac{s}{2} {\rm Tr} [ \mathbf{G}_{1} \mathbf{V}  \mathbf{G}_{1} \mathbf{V}  ] 
		\nonumber
		\\
		&= & 
		\frac{1}{2} \int_Q \Bigl[ \Pi_0^{ AA } ( Q ) A_{-Q} A_Q
		+ \Pi_0^{BB} ( Q )  B_{-Q } B_Q \Bigr],
		\hspace{7mm}
	\end{eqnarray}
	where
	\begin{subequations}
		\begin{eqnarray}
			\Pi_0^{AA} ( \bd{q} , i \nu  ) & = & \Pi_0^{ \bar{V} V } ( Q ) + \Pi_0^{ VV} ( Q) ,
			\\
			\Pi_0^{BB} ( \bd{q} , i \nu  ) & = & 
			\Pi_0^{ \bar{V} V } ( Q ) - \Pi_0^{ VV} ( Q).
			\hspace{9mm}
		\end{eqnarray}
	\end{subequations} 
	For simplicity let us assume that $\gamma_{ 2 \bd{k}_F + \bd{q}  \lambda} = \gamma_{\lambda}$
	and $\omega_{ 2 \bd{k}_F + \bd{q} \lambda} = \omega_{\lambda}$ are independent of the relative momentum
	$\bd{q}$ so that $V_{{Q}} = \sum_{\lambda} \gamma_{\lambda} V_{Q \lambda}$. Both approximations are justified for $\Delta \ll \epsilon_F$ where the physics is dominated by excitations around the Fermi surface.
	Similarly to $V_Q$, we expand $A_Q = \sum_{\lambda} \gamma_{\lambda} A_{Q \lambda}$ and
	$B_Q = \sum_{\lambda} \gamma_{\lambda} B_{Q \lambda}$ so that
	$V_{ Q \lambda} = \frac{1}{\sqrt{2}} [ A_{Q \lambda } + i B_{Q \lambda } ]$.
	Our Gaussian effective action (\ref{eq:SeffuV}) can then be written as
	\begin{eqnarray}
		& & S_{\rm eff} [  U , A, B  ]  =  \beta \Omega_1 + \beta {\cal{V}} \sum_{\lambda} \omega_{  \lambda}^2 
		| X^{0}_{  \lambda}  |^2 
		\nonumber
		\\
		&  & + \frac{1}{2} \int_{ Q} \sum_{\lambda}
		( \nu^2 + \omega^2_{\bd{q} \lambda } ) U_{ - Q \lambda } U_{ Q \lambda }
		\nonumber
		\\
		&  & + \frac{1}{2} \int_{ Q} \sum_{\lambda \lambda^{\prime}} 
		\Bigl\{
		[  \delta_{\lambda \lambda^{\prime}} ( \nu^2 + \omega^2_{ \lambda } )  + \gamma_{\lambda}
		\gamma_{\lambda^{\prime}} \Pi_0^{AA} ( Q ) ] A_{ -Q \lambda} A_{Q \lambda^\prime}
		\nonumber
		\\
		&  &  \hspace{10mm} +  [  \delta_{\lambda \lambda^{\prime}} ( \nu^2 + \omega^2_{ \lambda } )  + \gamma_{\lambda}
		\gamma_{\lambda^{\prime}} \Pi_0^{BB} ( Q ) ] B_{ -Q \lambda} B_{Q \lambda^{\prime}}    
		\Bigr\} ,
		\nonumber
		\\
		& &
		\label{eq:SeffuV2}
	\end{eqnarray}
	where in the first line we have introduced the 
	abbreviations $X^0_{\lambda} = X^0_{ 2 \bd{k}_F , \lambda}$.
	
	The collective modes in the CDW state can be identified with the eigenvectors of the matrices $\mathbf{M}^A ( Q)$ and $\mathbf{M}^B ( Q )$ in flavor space with matrix elements
	\begin{subequations}
		\begin{eqnarray}
			M^A_{\lambda \lambda^{\prime} }( Q ) & = & \delta_{\lambda \lambda^{\prime}} ( \nu^2 + \omega^2_{ \lambda } )  + \gamma_{\lambda}
			\gamma_{\lambda^{\prime}} \Pi_0^{AA} ( Q ),
			\hspace{7mm}
			\\
			M^B_{\lambda \lambda^{\prime} }( Q ) & = & \delta_{\lambda \lambda^{\prime}} ( \nu^2 + \omega^2_{ \lambda } )  + \gamma_{\lambda}
			\gamma_{\lambda^{\prime}} \Pi_0^{BB} ( Q ).
		\end{eqnarray}
	\end{subequations}
	The energy dispersions of the collective modes  can be obtained from the solutions of the equations
	\begin{subequations}
		\label{eq:detzero}
		\begin{eqnarray} 
			{\rm det} \mathbf{M}^A ( \bd{q} , \omega + i 0 ) & = & 0,
			\\
			{\rm det} \mathbf{M}^B ( \bd{q} , \omega + i 0 ) & = & 0,
		\end{eqnarray}
	\end{subequations}
	Introducing the diagonal matrix $\mathbf{D}$ with matrix elements
	\begin{equation}
		D_{\lambda \lambda^{\prime}} = \delta_{\lambda \lambda^{\prime}} ( \nu^2 + \omega_{\lambda}^2 ),
	\end{equation}
	and the column vector $\bd{\gamma}$ with components $\gamma_{\lambda}$,
	the above matrices $\mathbf{M}^A$ and $\mathbf{M}^B$ have the structure
	\begin{equation}
		\mathbf{M} = \mathbf{D} + \Pi_0 \bd{\gamma} \bd{\gamma}^T.
	\end{equation}
	Anticipating that for the relevant frequencies ${\rm det} \mathbf{D} \neq 0$ we find that
	the condition ${\rm det} \mathbf{M} =0$  reduces to
	\begin{equation}
		( \bd{\gamma}^T \mathbf{D}^{-1} \bd{\gamma} )^{-1} + \Pi_0 =0.
	\end{equation}
	We conclude that the eigenfrequencies of the amplitude modes 
	can be obtained from the roots of the equation
	\begin{equation}
		\left[ \sum_{\lambda} \frac{ \gamma_{\lambda}^2}{ \omega_{\lambda}^2 - \omega^2 } \right]^{-1}
		+ \Pi_0^{AA} ( \bd{q} , \omega + i 0 ) =0,
		\label{eq:eigenA}
	\end{equation}
	while the eigenfrequencies of the phase modes satisfy
	\begin{equation}
		\left[ \sum_{\lambda} \frac{ \gamma_{\lambda}^2}{ \omega_{\lambda}^2 - \omega^2 } \right]^{-1}
		+ \Pi_0^{BB} ( \bd{q} , \omega + i 0 ) =0.
		\label{eq:eigenB}
	\end{equation}
	Alternatively, these conditions can be obtained from the
	effective action $S_{\rm eff} [ A , B ]$
	of the average fields $A = \sum_{\lambda} \gamma_{\lambda} A_{\lambda}$ and
	$B = \sum_{\lambda} \gamma_{\lambda} B_{\lambda}$ defined by
	\begin{eqnarray}
		e^{ - S_{\rm eff} [ A , B ] } & = &  \int {\cal{D}} [ U ]
		\int {\cal{D}} [ A_{\lambda} ]  \int {\cal{D}} [ B_{\lambda} ] 
		e^{ - S_{\rm eff} [ U , A_{\lambda} , B_{\lambda} ]}
		\nonumber
		\\
		&  \times & \delta ( A - \sum_{\lambda} \gamma_{\lambda} A_{\lambda} )
		\delta ( B - \sum_{\lambda} \gamma_{\lambda} B_{\lambda} ) .
		\label{eq:HSaverage}
	\end{eqnarray}
	Representing the $\delta$-functions via functional integrals over auxiliary fields $\phi^A$ and $\phi^B$, carrying out the Gaussian integrations over $A_{\lambda}$ and $B_{\lambda}$ and then over
	$\phi^A$ and $\phi^B$, and dropping an additive  constant we obtain
	\begin{eqnarray}
		S_{\rm eff} [ A , B ] & = & \frac{1}{2} \int_Q \Bigl\{
		\left[  \bar{D}_0^{-1} ( 2 k_F , i \nu ) + \Pi_0^{AA} ( Q ) \right] A_{ - Q } A_Q 
		\nonumber
		\\
		&  & +  
		\left[   \bar{D}_0^{-1} ( 2 k_F , i \nu )   + \Pi_0^{BB} ( Q ) \right] B_{ - Q } B_Q \Bigr\}, \nonumber
		\\
		& &
		\label{eq:SeffAB}
	\end{eqnarray}
	where 
	\begin{equation}
		\bar{D}_0 ( 2 k_F, i \nu ) =  \sum_{\lambda} \frac{ \gamma_{\lambda}^2}{ \nu^2 + \omega_{\lambda}^2 }.
	\end{equation}
	We obtain  Eqs.~(\ref{eq:eigenA}) and (\ref{eq:eigenB}) from the condition
	that the corresponding Gaussian propagators have poles on the real frequency axis.
	
	To write the mode-frequency equations~(\ref{eq:eigenA}) and (\ref{eq:eigenB}) in a compact form we note that the generalized polarization functions $\Pi_0^{AA} ( \bd{q} , \omega + i0 )$ and $\Pi_0^{BB} ( \bd{q} , \omega + i0)$ appearing in these equations can both be expressed in terms of the following auxiliary function
	\begin{equation}\label{eq:deff}
		F ( \bd{q} ,i\nu)\equiv\frac{s}{2}\int_{K}\frac{1}{\big[(i\omega)^2-
			E_{\bd{k}}^2\big]\big[(i\omega+i\nu)^2-E^2_{\bd{k} + \bd{q}} \big]}.
	\end{equation}
	After carrying out the frequency sum and setting $\bd{q}=0$ this reduces to
	\begin{equation}
		F ( 0 ,i\nu) = \frac{s}{2 {\cal{V}} }\sum_{\bd{k}} 
		\frac{\tanh\left(\frac{\beta E_{\bd{k}}}{2}\right)}{E_{\bd{k}}\big[(2E_{\bd{k}})^2 + \nu^2\big]}.
		\label{eq:deffnum}
	\end{equation}
	Using the fact that at zero temperature
	the gap equation~\eqref{eq:selfcon} 
	implies 
	
	\begin{equation}
		\frac{s}{\cal{V}} \sum_{\bd{k}} \frac{1}{2 E_{\bd{k}}} = \left[ \sum_{\lambda} \frac{ \gamma^2_{\lambda}}{ \omega^2_{\lambda}} \right]^{-1}
		= \bar{D}^{-1}_0 ( 2 k_F , 0 ) ,
	\end{equation}
	we find, after analytic continuation to real frequencies, that Eq.~(\ref{eq:eigenA}) reduces to	
	\begin{eqnarray}
		& & \bar{D}_0^{-1} ( 2 k_F , \omega )-\bar{D}_0^{-1} ( 2 k_F , 0 ) \nonumber
		\\  &= & \Big[\omega^2 -  (v_Fq_x)^2 -  (2\Delta)^2\Big] F ( \bd{q} , \omega) .
		\label{eq:eigenA2}
		\hspace{7mm}
	\end{eqnarray}
	Similarly, we obtain from Eq.~(\ref{eq:eigenB}) for the frequencies of the phase modes
	\begin{equation}
		\bar{D}_0^{-1} ( 2 k_F , \omega )-\bar{D}_0^{-1} ( 2 k_F , 0 )
		=  \Big[\omega^2 -  (v_Fq_x)^2\Big]F ( \bd{q} , \omega).\label{eq:eigenB2}
	\end{equation}
	
	For simplicity, let us now focus on the zero temperature limit. For small frequencies and momenta $\bd{q} = q_x \bd{e}_x$ we may then approximate 
	
	\begin{equation}
		F(q_x, \omega)\approx F_0 \Bigg[1+  \frac{2}{3}\frac{ \omega^2 }{(2 \Delta )^2}-  \frac{2}{3} \frac{  (v_F q_x)^2}{(2\Delta)^2} \Bigg],
		\label{eq:Iexpansion}
	\end{equation}
	with
	\begin{equation}
		F_0 = \frac{ \nu_3}{2 (2 \Delta )^2},
	\end{equation}
	where the density of states $\nu_3$ is  defined in Eq.~(\ref{eq:nu3def}). Substituting the expansion (\ref{eq:Iexpansion}) into the mode-frequency equation (\ref{eq:eigenA2}) for the amplitude modes we obtain	
	
	\begin{eqnarray}
		& & 	\bar{D}_0^{-1} ( 2 k_F , \omega  )-\bar{D}_0^{-1} ( 2 k_F , 0 ) \nonumber
		\\ 
		& = &     \left[   \frac{\omega^2}{3} -   \frac{ ( v_F q_x )^2  }{3} -   (2\Delta)^2 \right] F_0 .
		\label{eq:ampliphon}
	\end{eqnarray}
	Anticipating that for K$_{0.3}$MoO$_3$ the solutions of this equation 
	are only perturbatively shifted from the bare phonon frequencies $\omega_{\lambda}$, we may approximate the  inverse phonon propagator in Eq.~(\ref{eq:ampliphon}) in the regime  $ | \omega - \omega_{\lambda} | \ll |\omega_{  \lambda}-\omega_{  \lambda\pm1}|$ by
	\begin{equation}
		\bar{D}_0^{-1} ( 2 k_F ,  \omega  )\approx \frac{\omega_{  \lambda}^2-\omega^2}{\gamma^2_\lambda}.
	\end{equation}
	With this approximation, we obtain from Eq.~(\ref{eq:ampliphon}) the following explicit expression for the squared frequency of the amplitude mode adiabatically related to the phonon mode $\lambda$,
	\begin{equation}
		\omega^2_{A,\lambda}(q_x) = \omega^2_{A, \lambda} (0) + ( v_{ A , \lambda} q_x )^2,
		\label{eq:omegaAres}
	\end{equation}
	with
	\begin{eqnarray}
		\omega^2_{A,\lambda}(0) & = & 
		\frac{\omega_{  \lambda}^2}{1+\frac{\gamma_{\lambda}^2 F_0 }{3}}
		\left[ 1 - \frac{ \gamma_{\lambda}^2}{\omega_{\lambda}^2}
		\bar{D}_0^{-1} ( 2 k_F , 0 )
		+  \frac{\nu_3 \gamma_{\lambda}^2  }{2 \omega_{\lambda}^2} \right],
		\nonumber
		\\
		& &
		\label{omegaAres}
		\\
		v_{A , \lambda}^2  & = & \frac{ \frac{\gamma_{\lambda}^2 F_0}{3} }{
			1 + \frac{ \gamma_{\lambda}^2 F_0}{3} } v_F^2.
	\end{eqnarray}
	
	Next, let us examine the equation  \eqref{eq:eigenB2} for the frequencies of the phase modes. This  equation always has a special low-energy solution with linear dispersion $\omega  \approx v_{\varphi} | q_x |$ for $ | q_x | \rightarrow0$ which can be identified with the gapless Goldstone mode implied by the spontaneous breaking of the $U(1)$ symmetry associated with the CDW order parameter. To see this, we note that for sufficiently  small $ \omega$ we may expand the left-hand side of Eq.~\eqref{eq:eigenB2} to quadratic order in $\omega$, 
	\begin{equation}
		\bar{D}_0^{-1} ( 2 k_F , \omega )-\bar{D}_0^{-1} ( 2 k_F , 0 )\approx- \omega^2 \frac{  \sum_{\lambda} \frac{
				\gamma_{\lambda}^2}{ \omega_{\lambda}^4 }  }{
			\left[  \sum_{\lambda}  \frac{ \gamma_{\lambda}^2}{ \omega_{\lambda}^2 } \right]^2 },
	\end{equation}
	which is sufficient to obtain the dispersion of the lowest-frequency phase mode for  
	small $ | q_x |$. Then Eq.~(\ref{eq:eigenB2}) reduces to
	\begin{equation}
		- \omega^2\left[ \frac{  \sum_{\lambda} \frac{
				\gamma_{\lambda}^2}{ \omega_{\lambda}^4 }  }{
			\left[  \sum_{\lambda}  \frac{ \gamma_{\lambda}^2}{ \omega_{\lambda}^2 } \right]^2 }+F_0\right]+F_0(v_Fq_x)^2=0,
	\end{equation}
	which implies for the squared frequency of the lowest phase mode
	\begin{equation}
		\omega^2_{B}(q_x)= ( v_\varphi  q_x  )^2.
	\end{equation}
	The squared phase velocity can be written as
	\begin{equation}
		v_\varphi^2 =   \frac{ g_1 }{ 1 + g_1} v_F^2,
		\label{eq:vpdef}
	\end{equation}
	where we have introduced the dimensionless coupling constant
	\begin{equation}\label{eq:omegaBres}
		g_1 =F_0	\frac{\left[  \sum_{\lambda}  \frac{ \gamma_{\lambda}^2}{ \omega_{\lambda}^2 } \right]^2 }{
			\sum_{\lambda} \frac{ \gamma_{\lambda}^2}{ \omega_{\lambda}^4 } 
		} =\frac{ \nu_3 }{ 2 (2 \Delta)^2 } 
		\frac{\left[  \sum_{\lambda}  \frac{ \gamma_{\lambda}^2}{ \omega_{\lambda}^2 } \right]^2 }{
			\sum_{\lambda} \frac{ \gamma_{\lambda}^2}{ \omega_{\lambda}^4 }}.
	\end{equation}
	
\begin{figure}[tb]
		\begin{center}
			\centering
			\includegraphics[width=0.45\textwidth]{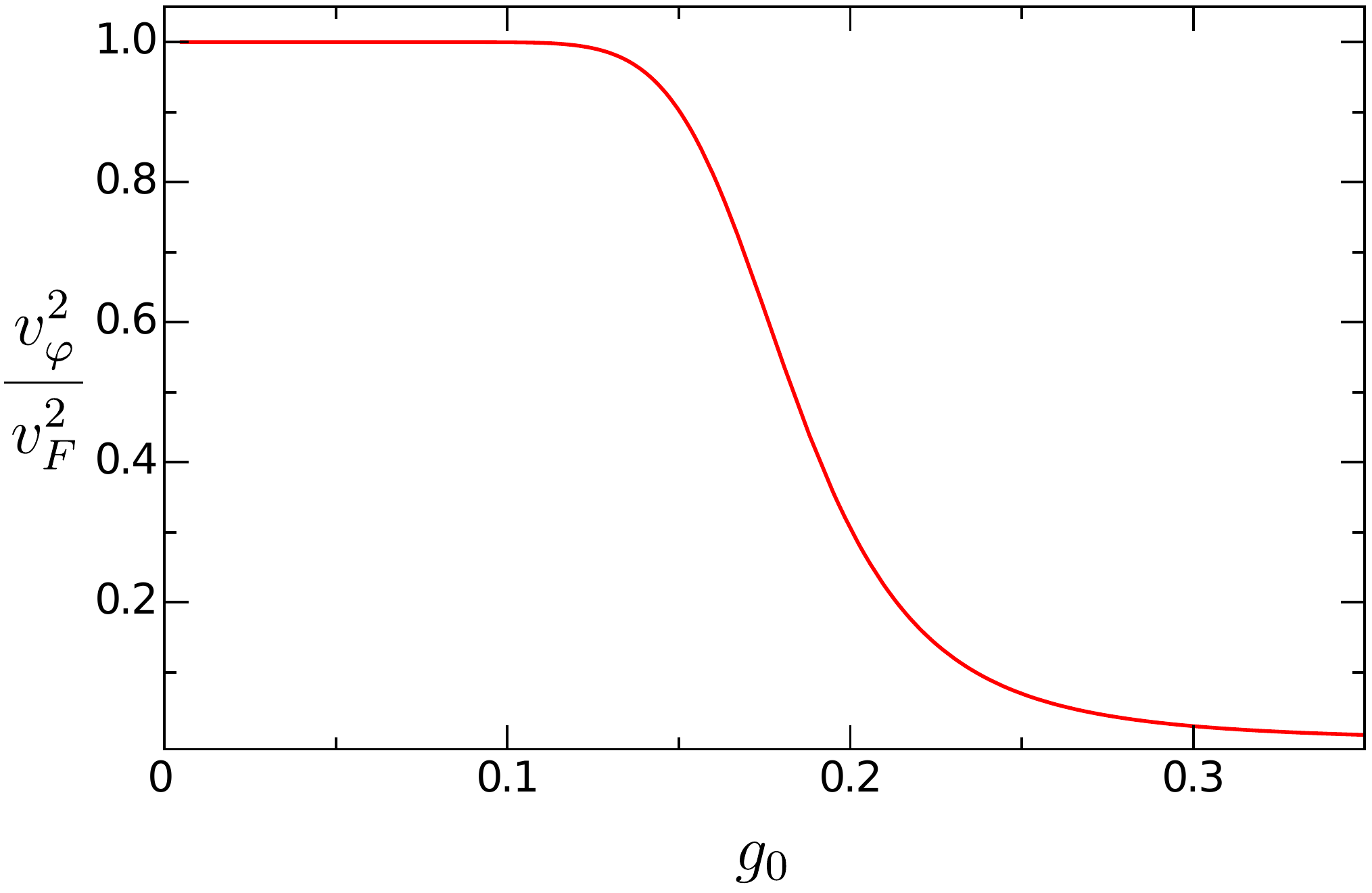}
		\end{center}
		\caption{
			Squared phase velocity  $ v_\varphi^2$ 
			given in Eq.~(\ref{eq:vpdef})
			in units of the squared Fermi velocity $v_F^2$  as a function of the
			dimensionless electron-phonon coupling $g_0$
			defined in Eq.~(\ref{eq:gnull}). The plot is for  $\omega_c / \omega_{ 2 \bd{k}_F} = 25$, where $\omega_c$ is the ultraviolet cutoff in the self-consistency 
			equation (\ref{eq:gapequation}) for the mean-field order parameter $\Delta$.}
		\label{fig:ZB}
	\end{figure}

	Finally, let us give the dispersions of the amplitude and phase modes in the special case where the system supports only a single phonon mode. Then we may omit the summations over the phonon label $\lambda$ so that our dimensionless coupling $g_1$ reduces to
	\begin{equation}
		g_1 = \gamma_{ 2 \bd{k}_F}^2 F_0 = g_0 \frac{ \omega_{ 2 \bd{k}_F}^2 }{ 4 \Delta^2}, 
		\label{eq:g1defsingle}
	\end{equation}
	where
	\begin{equation}\label{eq:gnull}
		g_0 =
		\frac{\nu_3 \gamma_{ 2 \bd{k}_F }^2}{2 \omega_{ 2 \bd{k}_F}^2}
	\end{equation}
	is the single-phonon version of the dimensionless electron-phonon coupling defined  in
	Eq.~(\ref{eq:g0def}).
	Our result  for the squared frequency of the amplitude mode in  (\ref{eq:omegaAres}) then reduces to
	\begin{equation}
		\omega_{ A}^2(q_x)=\frac{g_1}{1+\frac{g_1}{3} }\Big[(2\Delta)^2+\frac{(v_Fq_x)^2}{3}\Big],
	\end{equation}
	while the phason velocity in Eq.~(\ref{eq:vpdef}) can be written as
	\begin{equation}
		\frac{ v_\varphi^2}{v_F^2} = \frac{ g_0   }{  g_0     + 
			4 \Delta^2/  \omega_{ 2 \bd{k}_F}^2  }.
		\label{eq:vphivF}
	\end{equation}
	Note that for K$_{0.3}$MoO$_3$ we estimate $ 2 \Delta / \omega_{ 2 \bd{k}_F } \approx 12$ at $T=0$, so that $v_{\varphi}^2 / v_F^2 \approx g_0 \omega^2_{ 2 \bd{k}_F} /(4 \Delta^2 ) = g_1$, in agreement with the expression given in the review  by Gr\"{u}ner \cite{Gruener88,footnoteGruener}. A graph of the squared phase velocity in Eq.~(\ref{eq:vphivF}) as a function of the dimensionless electron-phonon coupling $g_0$ is shown in Fig.~\ref{fig:ZB}. 
	In general, the phason velocity is smaller than the Fermi velocity.  Though 
	in the regime of weak electron-phonon coupling, where $\Delta$ is exponentially small, the approximation
	$v_\varphi \approx v_F$ is very accurate.

	\section{Effect of long-range Coulomb interactions on the 
		collective modes}
	\label{sec:Coulomb}

	How are the above results modified by the long-range Coulomb interaction? According to Virosztek and Maki \cite{Virosztek93}, in the presence of long-range Coulomb interaction an incommensurate CDW at zero temperature does not have a	gapless collective mode related to the breaking of the $U(1)$-symmetry of the complex CDW order parameter at $T=0$. This is rather surprising, because it means that the Coulomb interaction completely destroys the Goldstone mode associated with the spontaneous breaking of the U(1)-symmetry of the complex order-parameter of the incommensurate CDW.  We show in this section that this result is erroneous and that the phase mode remains gapless even in the presence of Coulomb interactions. The crucial point is that the charge fluctuations associated with long-wavelength acoustic phonons screen the long-range Coulomb interaction even at zero temperature so that the resulting effective interaction is short-range and cannot qualitatively modify the gapless phase mode. This screening effect was not properly taken into account in Ref.~[\onlinecite{Virosztek93}].

	Starting point of our investigation is the Hamiltonian ${\cal{H}}_{\rm FC}$ given in
	Eq.~(\ref{eq:FroehlichC}) which is obtained by adding 	the  quantized	Coulomb energy of electronic and ionic charge fluctuations
	to the 	Fr\"{o}hlich Hamiltonian for generic electron-phonon interactions (\ref{eq:Froehlich}).
	After integrating the corresponding Euclidean action over the phonon momenta
	and decoupling the Coulomb interaction employing a Hubbard-Stratonovich field $\phi$ we obtain
	the following Euclidean action of our Fr\"{o}hlich-Coulomb model,
	\begin{eqnarray}
		S_{\rm FC} & = & \int_{ K \sigma} ( \epsilon_{\bd{k}} - \mu - i \omega ) \bar{c}_{ K \sigma} c_{ K \sigma}
		\nonumber
		\\
		& + & \frac{1}{2} \int_{ Q \lambda} ( \nu^2 + \omega_{\bd{q} \lambda}^2 ) X_{ - Q \lambda } X_{ Q \lambda} 
		\nonumber
		\\
		& + &  \frac{1}{2} \int_Q f_{\bd{q}}^{-1} \phi_{ - Q } \phi_Q 
		\nonumber
		\\
		& + & \int_{ K \sigma} \int_{ Q \lambda} \gamma_{\bd{q} \lambda}
		\bar{c}_{ K + Q \sigma} c_{ K \sigma} X_{ Q \lambda}
		\nonumber
		\\
		& + & i \int_Q \phi_{ - Q } \left[
		\int_{K \sigma} \bar{c}_{K + Q \sigma} c_{ K \sigma} + \sum_{\lambda} \alpha_{\bd{q} \lambda} X_{ Q \lambda} \right].
		\nonumber
		\\
		& &
	\end{eqnarray}
	In the last line the term  $i \int_Q \phi_{ - Q } \sum_{\lambda} \alpha_{\bd{q} \lambda} X_{ Q \lambda }$ 
	describes the contribution of  ionic charge fluctuations to the effective interaction. 
	It turns out that this term, which has been ignored 
	by Virosztek and Maki \cite{Virosztek93}, is essential to correctly describe the effect of Coulomb interactions on the phase modes in a CDW.
	
	Note that  pure Coulomb interactions in one-dimensional electronic systems  are generally non-perturbative already at low energies due to the Luttinger-liquid physics  \cite{Giamarchi_Book} dominated by the gapless CDW states, and a growing body of evidence indicates that the formation of CDWs  in these systems is more generic then the low-energy phenomenon~\cite{Vianez22,Tsyplyatyev22}. Nevertheless, the use of only the Gaussian approximation is justified in our calculation by the effect of
	the electron-phonon interaction which triggers the 
	formation of the gapful correlated CDW state via the Peierls instability. At low temperatures  
	the inverse of the large value of the  order parameter $\Delta$ 
	in Eq.~\eqref{Delta_order_parameter}  
	provides  a small parameter for the perturbative treatment of 
	the electron-electron interaction even in one dimension in this work. Also some numerical studies of the interplay between Luttinger liquids and the Peierls instability in \cite{Fradkin83,Weiss98} suggest a possibility of such a scenario. We will return to this point later in Sec.~\ref{sec:experiment} where we compare the results of this section with the experimental data. 
	
	Following the procedure outlined in Sec.~\ref{sec:modesF} we may now integrate over the
	fermions in the two-band approximation to obtain the effective action of the	phonons and the Coulomb field,
	\begin{eqnarray}
		S_{\rm eff} [ X , \phi ] & = &  \frac{1}{2} \int_{ Q \lambda} ( \nu^2 + \omega_{\bd{q} \lambda}^2 ) X_{ - Q \lambda } X_{ Q \lambda} 
		\nonumber
		\\
		& + &  \frac{1}{2} \int_Q f_{\bd{q}}^{-1} \phi_{ - Q } \phi_Q + i \int_Q \phi_{ - Q } U_{Q}
		\nonumber
		\\
		& - & s {\rm Tr} \ln ( \mathbf{G}_1^{-1} - \mathbf{V} ),
		\label{eq:SeffXphi}
	\end{eqnarray}
	where the electronic propagator $\mathbf{G}_1$ in mean-field approximation
	is defined in Eq.~(\ref{eq:G1def}) and the fluctuation matrix 
	$\mathbf{V} $ is 
	now given by
	\begin{eqnarray}
		{[} \mathbf{V} {]}_{ K K^{\prime}} & = & 
		\left( \begin{array}{cc}  i \phi_{ K - K^{\prime}}  & 
			V_{ K - K^{\prime}} \\
			V^{\ast}_{ K^{\prime} -K} &  i \phi_{ K - K^{\prime}} \end{array} \right).
	\end{eqnarray}
	Here the  low-energy fields $U_{Q}    $ and $X_{Q \lambda}$ are defined in Eq.~(\ref{eq:UVlow}), i.e.,
	$U_{ Q \lambda } = X_{ Q \lambda}$ for $ | \bd{q} | \ll k_F$ and
	$V_{Q \lambda} = X_{ 2 \bd{k}_F + Q \lambda} - \delta_{ Q ,0 }  X^{0 }_{\lambda}$,
	with
	$U_Q = \sum_{\lambda} \alpha_{ \bd{q} \lambda} U_{ Q \lambda}$
	and $V_Q = \sum_{\lambda} \gamma_{\lambda} V_{ Q \lambda}$.
	The next few steps are analogous to the steps
	in Sec.~\ref{sec:modesF} so that we relegate them to Appendix A. 
	Our final result for the effective action of the collective amplitude and phase modes in the presence of long-range Coulomb interactions is
	\begin{eqnarray}
		& & S_{\rm eff} [ A , B ]  =  
		\frac{1}{2} \int_Q \biggl\{ \left[ \bar{D}_0^{-1} ( 2 k_F , i \nu ) 
		+ \Pi_0^{AA} ( Q ) \right] A_{ -Q } A_Q 
		\nonumber
		\\
		&  & + \left[ \bar{D}_0^{-1} ( 2 k_F , i \nu ) 
		+ \Pi_0^{BB} ( Q ) + \Pi_{1}^{BB} ( Q ) \right] B_{ -Q } B_Q \biggr\},
		\nonumber
		\\
		& &
		\label{eq:SeffABCB}
	\end{eqnarray}
	where the generalized polarization function 
	$\Pi_1^{BB} ( Q )$ is given by
	\begin{equation}
		\Pi_1^{BB} ( Q ) = - \frac{2 f_{\bd{q}} \Pi_0^{ \phi V} ( -Q ) \Pi_0^{\phi V} ( Q ) }{ 1 + 
			f_{\bd{q}} [ \Pi_0^{\phi \phi} ( Q ) + \bar{D}_0 ( Q ) ] }.
		\label{eq:Pi1}
	\end{equation} 
	Here
	\begin{equation}
		\bar{D}_0 ( Q ) = \sum_{\lambda}
		\frac{ \alpha^2_{ \bd{q} \lambda}}{ \nu^2 + \omega^2_{\bd{q} \lambda} }
	\end{equation}
	is the propagator of long-wavelength phonons
	and the polarization functions
	$\Pi_0^{\phi V} ( Q )$ and
	$\Pi_0^{\phi \phi} ( Q )$ are defined in Eq.~(\ref{eq:pidefs2}).
	By comparing the effective action (\ref{eq:SeffABCB}) with the
	corresponding effective action (\ref{eq:SeffAB}) in the absence of Coulomb
	interactions, we see that within the Gaussian approximation the 
	amplitude modes are  not affected by the Coulomb interaction. On the other hand,
	the Coulomb interaction 
	gives rise to an additional contribution $\Pi_1^{BB} ( Q )$ to the
	polarization function of the phase modes.
	Note that the denominator in Eq.~(\ref{eq:Pi1}),
	\begin{eqnarray}
		\epsilon ( \bd{q} , i \nu ) & = &  
		1 + f_{\bd{q}} [ \Pi_0^{\phi \phi} ( Q ) + \bar{D}_0 ( Q ) ]
		\nonumber
		\\
		& = & 1  + f_{\bd{q}} \left[ \Pi_0^{\phi \phi} ( Q ) + \sum_{\lambda}		\frac{ \alpha^2_{ \bd{q} \lambda}}{ \nu^2 + \omega^2_{\bd{q} \lambda} }  \right],
		\label{eq:dielectricfunc}
	\end{eqnarray}
	can be identified with the long-wavelength dielectric function in the CDW state.
	After analytic continuation to real frequencies the second term in the square bracket
	is the usual contribution of harmonically bound charges to the dielectric 
	function~\cite{Jackson75}.

\begin{figure}[tb]
		\begin{center}
			\centering
			\vspace{7mm}
			\includegraphics[width=0.45\textwidth]{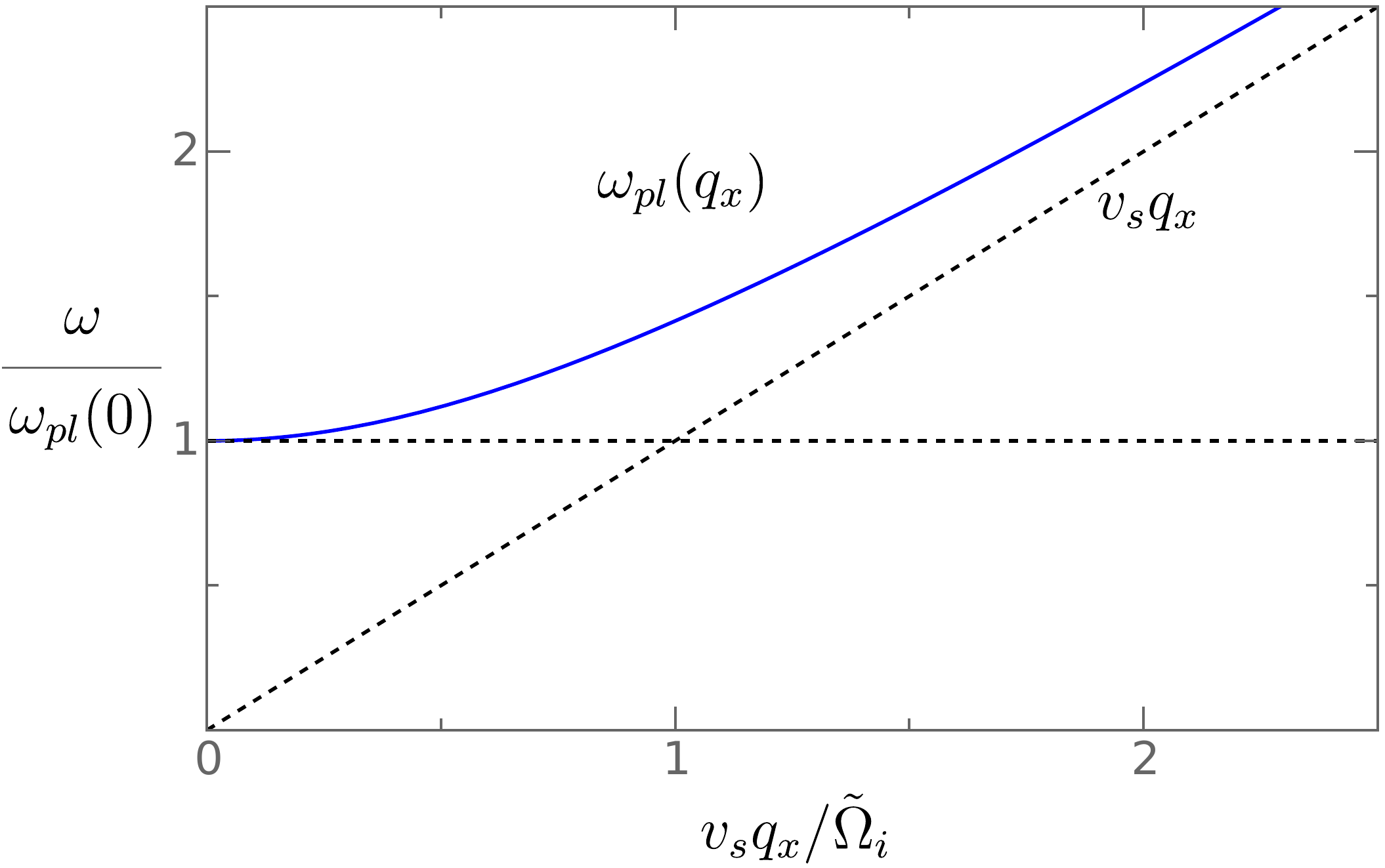}
		\end{center}
		\caption{
			Qualitative behavior of dispersion  $\omega_{\rm pl} (q_x)$ of plasma oscillations 
			in a CDW given in Eq.~(\ref{eq:Iqdef}). The crossover momentum  is given by $\tilde{\Omega}_i / v_s$ where $\tilde{\Omega}_i = \Omega_i / \sqrt{\epsilon}_r$ is the renormalized ionic plasma frequency, see
			Eq.~(\ref{eq:Omegaionren}).
		}
		\label{fig:plasma}
	\end{figure}

	To derive the dispersions of the long-wavelength phase modes
	we expand the polarization functions in Eq.~(\ref{eq:SeffABCB})
	for small  $ | v_F  q_x |\ll \omega_{  \lambda} \ll 2\Delta$ and $ | \nu | \ll 2\Delta$. At zero temperature we obtain  
	\begin{eqnarray}
		\Pi_0^{\phi \phi} ( \bd{q} , i \nu ) & \approx & \nu_3 \frac{ ( v_F q_x )^2}{6 \Delta^2 },  \label{eq:PiphiphiT0}
		\\
		\Pi_0^{\phi V} ( \bd{q} , i \nu ) & \approx &  \nu_3 \frac{ v_F q_x }{ 4 \Delta }.
	\end{eqnarray}
	For simplicity, let us now assume that the system supports only a single longitudinal acoustic phonon with linear dispersion 
	\begin{equation}
		\omega_{\bd{q}} = v_s | \bd{q} |,
	\end{equation}
	with the sound velocity $v_s$. Then we may omit the $\lambda$-summation
	in Eq.~(\ref{eq:dielectricfunc}) and the coupling $\alpha_{\bd{q}}$ is given in 
	Eq.~(\ref{eq:alphadef1}). Setting $\bd{q} = q_x \bd{e}_x$ and using the fact that according to
	Eq.~(\ref{eq:ionicplasma}) the combination $f_{\bd{q}} \alpha_{\bd{q}}^2 = \Omega_i^2$ can be identified with the square of the ionic plasma frequency, the dielectric function $\epsilon  ( q_x , i \nu )$ in Eq.~(\ref{eq:dielectricfunc}) can  be written as
	\begin{equation}
		\epsilon ( q_x , i \nu ) = 1 +  \frac{ \Omega_e^2}{ 6 \Delta^2}  + \frac{ \Omega_i^2}{ 
			(v_s q_x )^2 + \nu^2 },
	\end{equation}
	where the squares of the ionic and electronic plasma frequencies are
	\begin{eqnarray}
		\Omega_i^2 & = & \frac{ 4 \pi ( z e )^2 n_i}{M},
		\\
		\Omega_e^2 & = & \frac{ 4 \pi e^2 n}{m} = 4 \pi e^2 \nu_3 v_F^2.
	\end{eqnarray}
	Here $n_i$ is the density of ions, $n$ is the density of the electrons, and $m$ is the electronic mass.
	The energy dispersion $ \omega_{\rm pl} (  q_x ) $ of collective plasma oscillations can now be obtained from $ \epsilon ( q_x , \omega + i 0 )=0$, which gives
	\begin{equation}
		\omega_{{\rm pl}} ( q_x ) =  \sqrt{ \frac{ \Omega_i^2}{\epsilon_r} + (v_s q_x)^2},
		\label{eq:Iqdef}
	\end{equation}
	where
	\begin{equation}
		\epsilon_r  = 1 + \frac{ \Omega_e^2}{ 6 \Delta^2}
	\end{equation}
	is the static relative dielectric constant in the CDW state.
	The qualitative behavior of the frequency $\omega_{\rm pl} ( q_x )$ of plasma oscillations
	in a CDW is illustrated in Fig.~\ref{fig:plasma}. 
	Note that 
	for $ v_s | q_x | \ll \Omega_i/ \sqrt{\epsilon_r}$ and $\Delta \ll \Omega_e$, where
	$\epsilon_r \sim \Omega_e^2 / (6\Delta^2) \gg 1$,
	the scale of plasma oscillations in the
	one-dimensional CDW is given by 
	\begin{equation}
		\frac{\Omega_i }{ \sqrt{\epsilon_r}} \propto
		\frac{\Omega_i  } { \Omega_e }  \Delta  \propto \sqrt{ \frac{ m}{M}} 
		\Delta.
	\end{equation}
	On the other hand,
	in the opposite limit $ v_s | q_x | \gg \Omega_i/ \sqrt{\epsilon_r}$
	the plasmon oscillates with the frequency $v_s | q_x |$, i.e. in this regime the plasmon can be identified with the longitudinal acoustic phonon.
	
	Due to the coupling of the phase mode to the Coulomb field in the CDW state,
	the plasma oscillations hybridize with the phase modes.
	The energy dispersions of the hybrid modes can be obtained
	from the zeros of the inverse propagator of the phase modes in 
	our effective action $S_{\rm eff} [ A , B ]$ defined in Eq.~(\ref{eq:SeffABCB}),
	\begin{equation}
		\bar{D}_0^{-1} ( 2 k_F , \omega ) + \Pi_0^{BB} ( q_x , \omega )
		+ \Pi^{BB}_{1} ( q_x , \omega ) =0.
		\label{eq:hybridpp}
	\end{equation}
	Substituting the above long-wavelength limits for the generalized polarizations and focusing again on the single-phonon case, we find that Eq.~(\ref{eq:hybridpp}) can be reduced to a quadratic equation for the squared energies $ \omega^2_{\pm} ( q_x   )$ of the hybrid modes. The solutions of Eq.~(\ref{eq:hybridpp}) can be written as
	
	\begin{eqnarray}
		& &  \omega^2_{ \pm} ( q_x  )   =  \frac{1}{2} \left[
		\tilde{\Omega}_i^2 + \tilde{\Omega}_e^2  +  ( v_s^2 + v_\varphi^2) q_x^2  \right]
		\nonumber
		\\
		&  & \pm \sqrt{  \frac{1}{4} \left[ \tilde{\Omega}_i^2 - \tilde{\Omega}_e^2  
			+ ( v_s^2 - v_\varphi^2 ) q_x^2   
			\right]^2 +   \tilde{\Omega}_i^2 \tilde{\Omega}_e^2}  ,
		\hspace{7mm}
		\label{eq:omegapm}
	\end{eqnarray}
	where we have introduced renormalized ionic and electronic plasma frequencies
	\begin{eqnarray}
		\tilde{\Omega}_i^2 & = &  \frac{ \Omega_i^2}{\epsilon_r},
		\label{eq:Omegaionren}
		\\
		\tilde{\Omega}_e^2 & = & \frac{ v_\varphi^2}{v_F^2} \frac{ \Omega_e^2}{\epsilon_r} = \frac{ 4 \pi e^2 \nu_3 v_\varphi^2}{\epsilon_r },
	\end{eqnarray}
	and  $v_\varphi$ denotes the phason velocity in the absence of Coulomb interactions defined in Eq.~(\ref{eq:vpdef}).
	\begin{figure}[tb]
		\begin{center}
			\centering
			\vspace{7mm}
			\includegraphics[width=0.45\textwidth]{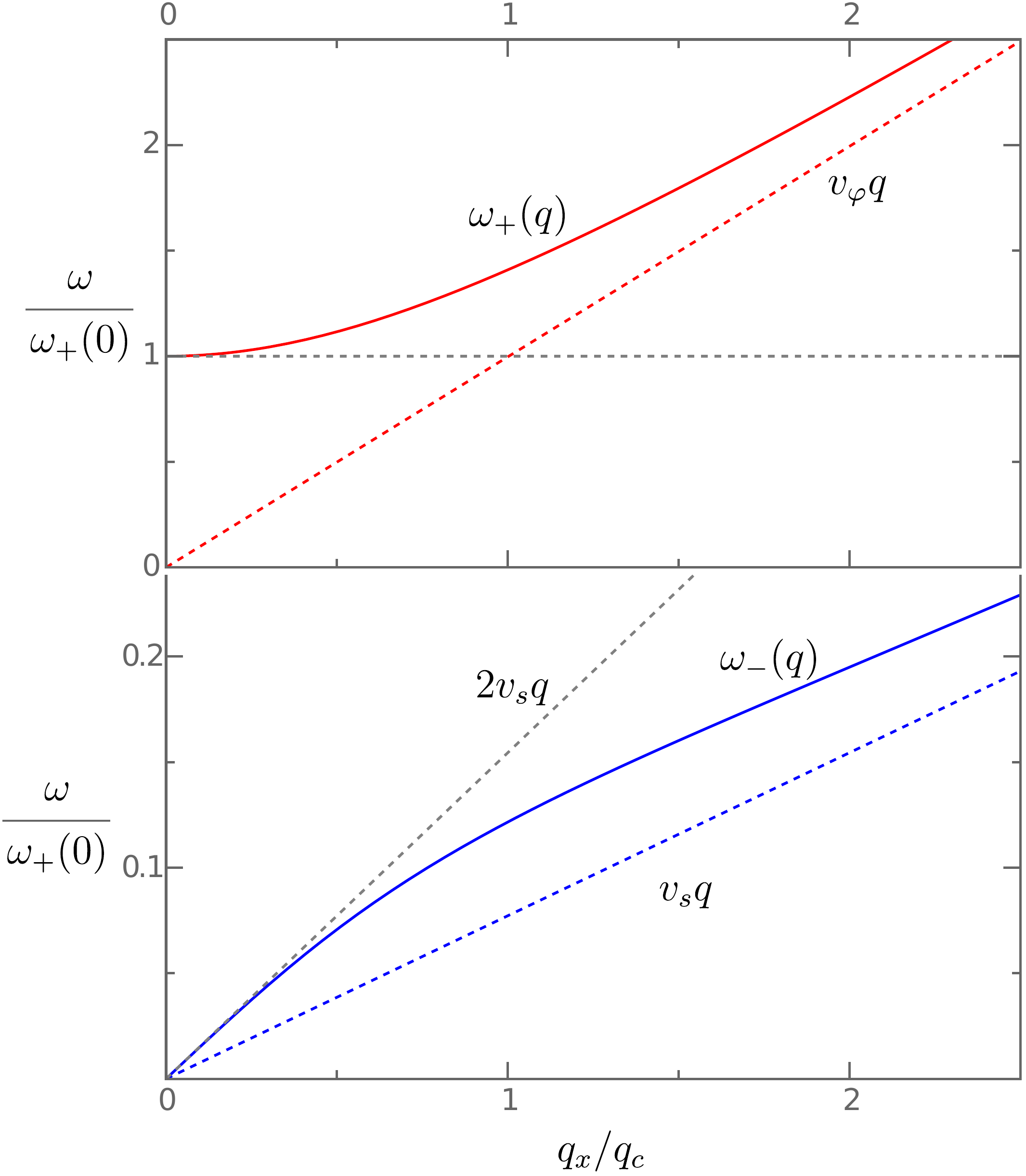}\\
		\end{center}
		\caption{
			Upper panel: Energy dispersion  of the gapped hybrid mode $\omega_{+} (  q_x )$ resulting from the hybridization of the plasmon with the phase modes given in Eq.~(\ref{eq:omegapm}). Lower panel: gapless hybrid mode  $\omega_{-} (  q_x )$. The plots are for $v_\varphi =  v_F $ and the crossover scale $q_c$
			is defined by the condition $\tilde{\Omega}_i^2 + \tilde{\Omega}_e^2 = ( v_s^2 + v_{\varphi}^2 ) q_c^2$.
		}
		\label{fig:hybridmode}
	\end{figure}
	A graph of the dispersions 
	$\omega_{\pm} ( q_x )$ in the regime where $v_{\varphi} \approx v_F$ is shown in  Fig.~\ref{fig:hybridmode}. 	Note that the lower mode $\omega_{-} ( q_x )$ has a gapless linear dispersion with velocity $2 v_s$ for small $|q_x |$. In fact, the asymptotic behavior for small $q_x$ can be obtained from Eq.~(\ref{eq:omegapm}),
	\begin{eqnarray}
		\omega^2_+ ( q_x ) & = &  \tilde{\Omega}_i^2 + \tilde{\Omega}_e^2  
		+  ( x_i v_s^2 + x_e v_\varphi^2 ) q_x^2 
		+ {\cal{O}} ( q_x^4 ), 
		\hspace{7mm}
		\\
		\omega^2_- ( q_x ) & = &
		( x_e v_s^2 + x_i v_\varphi^2 ) q_x^2 
		+ {\cal{O}} ( q_x^4),
	\end{eqnarray}
	where
	\begin{equation}
		x_i = \frac{ \tilde{\Omega}_i^2}{ \tilde{\Omega}_i^2 + \tilde{\Omega}_e^2 } , \; \; \;
		x_e = \frac{ \tilde{\Omega}_e^2}{ \tilde{\Omega}_i^2 + \tilde{\Omega}_e^2 }.
	\end{equation}

	\noindent In the weak coupling regime where  $2\Delta \ll \Omega_e$ the energy of the upper branch for $\bd{q}=0$ is  given by
	\begin{equation}\label{eq:omegaplusdelta}
		\omega_{+} ( 0 ) = \sqrt{\tilde{\Omega}_i^2 + \tilde{\Omega}_e^2}
		\approx \sqrt{6}\Delta\sqrt{\frac{\Omega_i^2}{\Omega_e^2}+\frac{v_\varphi^2}{v_F^2}}.
	\end{equation}
	On the other hand, the energy $\omega_- (q_x) $ of the lower branch 
	disperses linearly for small $q_x$,
	\begin{equation}
		\omega_- ( q_x ) \sim \tilde{v}_\varphi | q_x | ,
	\end{equation}
	with the renormalized phase velocity $\tilde{v}_\varphi$ given by
	\begin{equation} 
		\tilde{v}_\varphi =  \sqrt{ x_e
			v_s^2 + x_i v_\varphi^2 } .
		\label{eq:tildevB}
	\end{equation}
	Thus, in the presence of Coulomb interactions the velocity $\tilde{v}_\varphi$ of the gapless phase mode is a weighted average of the phonon velocity and the phason velocity without Coulomb interactions; the weights are determined by the ratio of the ionic and electronic plasma frequencies. In the adiabatic limit, where the mass $M$  of the ions is much larger than the mass $m$  of the electrons and the ionic plasma frequency is small compared with the electronic plasma frequency, we can  approximate the weights as $x_e \approx 1$ and $x_i \approx \tilde{\Omega}_i^2 / \tilde{\Omega}_e^2 	= (z m / M) v_F^2 / v_{\varphi}^2$. 
	\begin{figure}[b]
		\centering
		\includegraphics[width=1\linewidth]{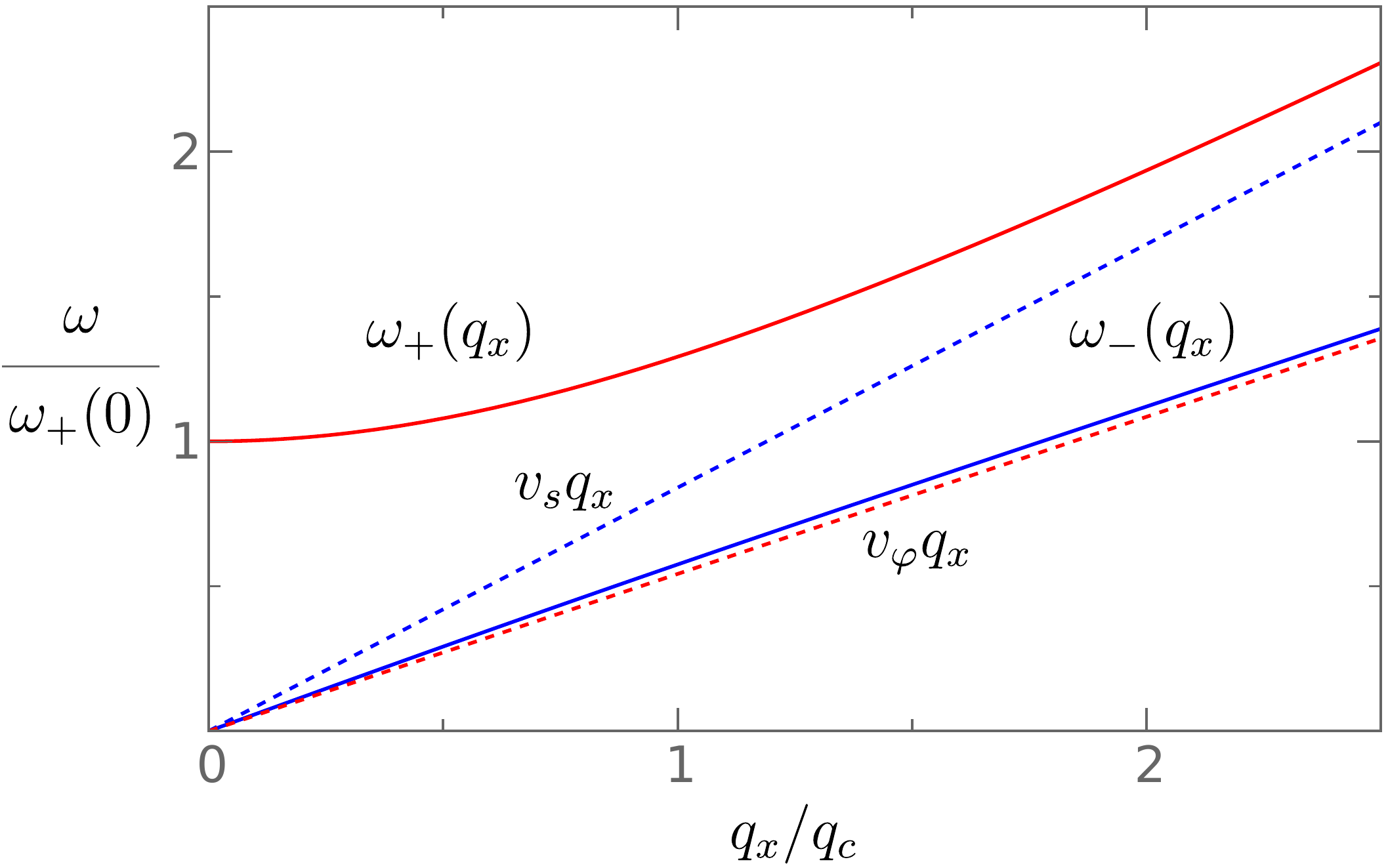}
		\caption{Energy dispersions  $\omega_{\pm} (  q_x )$ resulting from the	hybridization of the plasmons with the phase modes given in Eq.~(\ref{eq:omegapm}) in the limit of $v_\varphi\ll v_s\ll v_F$. The crossover scale $q_c$ is defined in the caption of Fig.~\ref{fig:hybridmode}.
		}
		\label{fig:real}
	\end{figure}
	Assuming in addition a weak electron-phonon coupling, i.e., $v_\varphi \approx v_F$,  we find that Eq.~(\ref{eq:tildevB}) reduces to
	\begin{equation}
		\tilde{v}_\varphi \approx  \sqrt{ v_s^2 + \frac{ \tilde{\Omega}_i^2}{\tilde{\Omega}_e^2} v_\varphi^2} 
		=  \sqrt{ v_s^2 + z \frac{m}{M} v_F^2 }.
	\end{equation}
	The Bohm-Staver relation \cite{Bohm51,Ashcroft76} allows us  	to express the sound velocity in terms of the Fermi velocity
	\begin{equation}
		v_s = \sqrt{ \frac{ z m}{3M}} v_F,
	\end{equation}
	which is justified because by definition $v_s$ is the bare sound velocity in the normal state of our model. Using the Bohm-Staver relation  we obtain in the adiabatic regime and for weak electron-phonon coupling
	\begin{equation}\label{eq:vstilde}
		\tilde{v}_{\varphi} = \sqrt{ \frac{ 4 z m}{3M}} v_F = 2 v_s.
	\end{equation}
	
	We conclude that  in this regime long-range Coulomb interactions can strongly renormalize the phason velocity in the ground state of a quasi one-dimensional CDW, as illustrated in the lower panel of Fig.~\ref{fig:hybridmode}: While $v_{\varphi} \approx v_F$ in the absence of Coulomb interactions, the inclusion of Coulomb interactions generates a strong renormalization of the phase velocity to a value of the order of the sound velocity $v_s$.

	Finally, consider the regime of intermediate to strong electron-phonon coupling where $v_{\varphi} \ll v_s \ll v_F$. In this regime, the coupling constant $g_1$ in Eq.~(\ref{eq:vpdef}) is small compared with unity so that $v_{\varphi} \ll v_F$ in the absence of Coulomb interactions. The renormalized value $\tilde{v}_{\varphi}$ of the phase velocity then has the same order of magnitude as the phase velocity  $v_{\varphi}$ without Coulomb interactions because the factors $x_i$ and $x_e$ in Eq.~(\ref{eq:tildevB}) have the same order of magnitude. The dispersions of the collective modes in this regime are illustrated in Fig.~\ref{fig:real}.
	We emphasize that the  gapless nature of the phase mode in the presence of long-range Coulomb interactions is protected by the existence of a longitudinal acoustic phonon in a crystal. This is an interesting example for the interplay of two Goldstone modes associated with the breaking of two different continuous symmetries, $U(1)$-symmetry of the CDW order parameter and translational symmetry of free space in a crystal.

	\section{Comparison with experiments}
	\label{sec:experiment}
	
	In this section, we fit the free parameters of our two model Hamiltonians  (with and without Coulomb interactions) using the temperature dependence of the lowest three amplitude and phase modes of K$_{0.3}$MoO$_3$ over a wide temperature range.	Then, we show that the predictions of our  calculations for the amplitude modes fit the available  experimental data \cite{Schaefer10,Schaefer14,Thomson17,Thomson23} remarkably well. However, for the phase modes, both models describe the experimental data given in Refs.~[\onlinecite{Thomson17,Thomson23}] only qualitatively. We, therefore,  discuss the shortcomings 		of our theoretical description of K$_{0.3}$MoO$_3$ and possible improvements to achieve a qualitatively accurate modeling of the experiments.

	\subsection{K$_{0.3}$MoO$_3$ and the multi-phonon Fr\"{o}hlich model for generic electron-phonon interactions}
	
	The multi-phonon Fr\"{o}hlich Hamiltonian ${\cal{H}}_F$ defined in Eq.~(\ref{eq:Froehlich}) with $\lambda_{\rm max}$ phonons depends on $2\lambda_{\rm max}$ free parameters: The phonon frequencies $\omega_{ 2 \bd{k}_F  \lambda}$ and the electron-phonon couplings $\gamma_{ 2 \bd{k}_F \lambda}$. In principle, these parameters can be determined by measuring the energies of the $\lambda_{\rm max}$ amplitude modes and the associated $\lambda_{\rm max}$ phase modes at some fixed momentum $\bd{q}$. Unfortunately, such a measurement is not available at this point,  since the experiments \cite{Schaefer10,Schaefer14,Thomson17,Thomson23} lack momentum resolution. Moreover, in contrast to our results for the  amplitude modes, our experimental data for the phase modes exhibit an unexplained temperature dependence  and no mode at zero frequency which prevents us from fitting our theoretical results to the raw data for the  $2\lambda_{\rm max}$ collective modes. Instead, we fix the $ 2 \lambda_{\rm max}$ free parameters of our multi-phonon Fr\"{o}hlich model using the frequencies $\omega_{\lambda} \equiv \omega_{ 2 \bd{k}_F \lambda}$ of the bare phonon modes from a measurement at $T>T_c$ and fitting the measured  amplitude modes at $T=0$, which can then be used to fix the squared electron-phonon couplings $\gamma_\lambda^2$. Remarkably, if we neglect the Coulomb interaction none of the subsequent findings are sensitive to the precise value of the gap $\Delta_0\equiv\Delta ( T=0)$ even though $\Delta_0$ is an additional free parameter. This is because the solutions to the mode equations~\eqref{eq:eigenA2} and \eqref{eq:eigenB2} are of the order of the phonon frequencies $\omega_{ 2 \bd{k}_F \lambda}$, which means that our calculation of the mode energies is  not only controlled by the dimensionless electron-phonon couplings $g_{0,\lambda}=\gamma_{ \lambda}^2\nu_3/2\omega_{ 2 \bd{k}_F  \lambda}^2$, but also by  the small parameter $\omega/2\Delta_0\sim\omega_{ 2 \bd{k}_F \lambda}/2\Delta_0$. Hence, the auxiliary function $F (0, \omega)$ in Eq.~\eqref{eq:deff} can be evaluated for small $\omega$ and, therefore, all corrections which explicitly depend on $\Delta_0$ are at least of order  $\omega^2/2\Delta_0^2 \ll 1$. \\
	
	\begin{figure}
		\centering
		\includegraphics[width=1\linewidth]{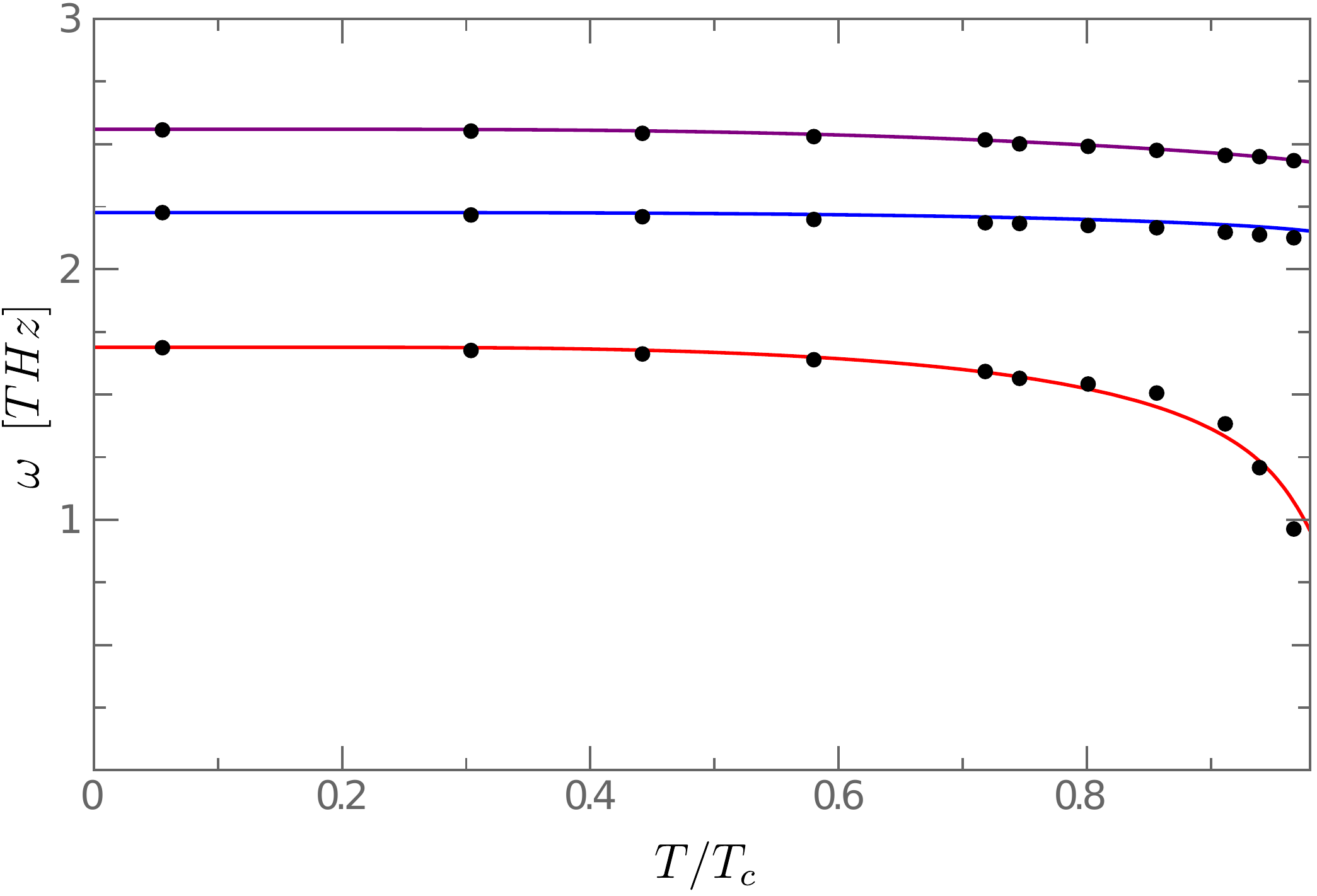}
		\caption{Lowest three amplitude modes of K$_{0.3}$MoO$_3$. The black dots are the measurements from  Refs.~\cite{Schaefer10,Schaefer14}, reanalyzed in \cite{Thomson23}, while the solid lines are the predictions of the amplitude mode equation Eq.~\eqref{eq:eigenA2} for $T>0$ and $q=0$, where we matched the coupling constants to the $T=0$ values taking the phonon frequencies from Refs.~\cite{Schaefer10,Schaefer14,Thomson17,Thomson23}, see Appendix B for the explicit values.}
		\label{fig:ExAmplidude}
	\end{figure}
	
\begin{figure}[tb]
		\centering
		\includegraphics[width=1\linewidth]{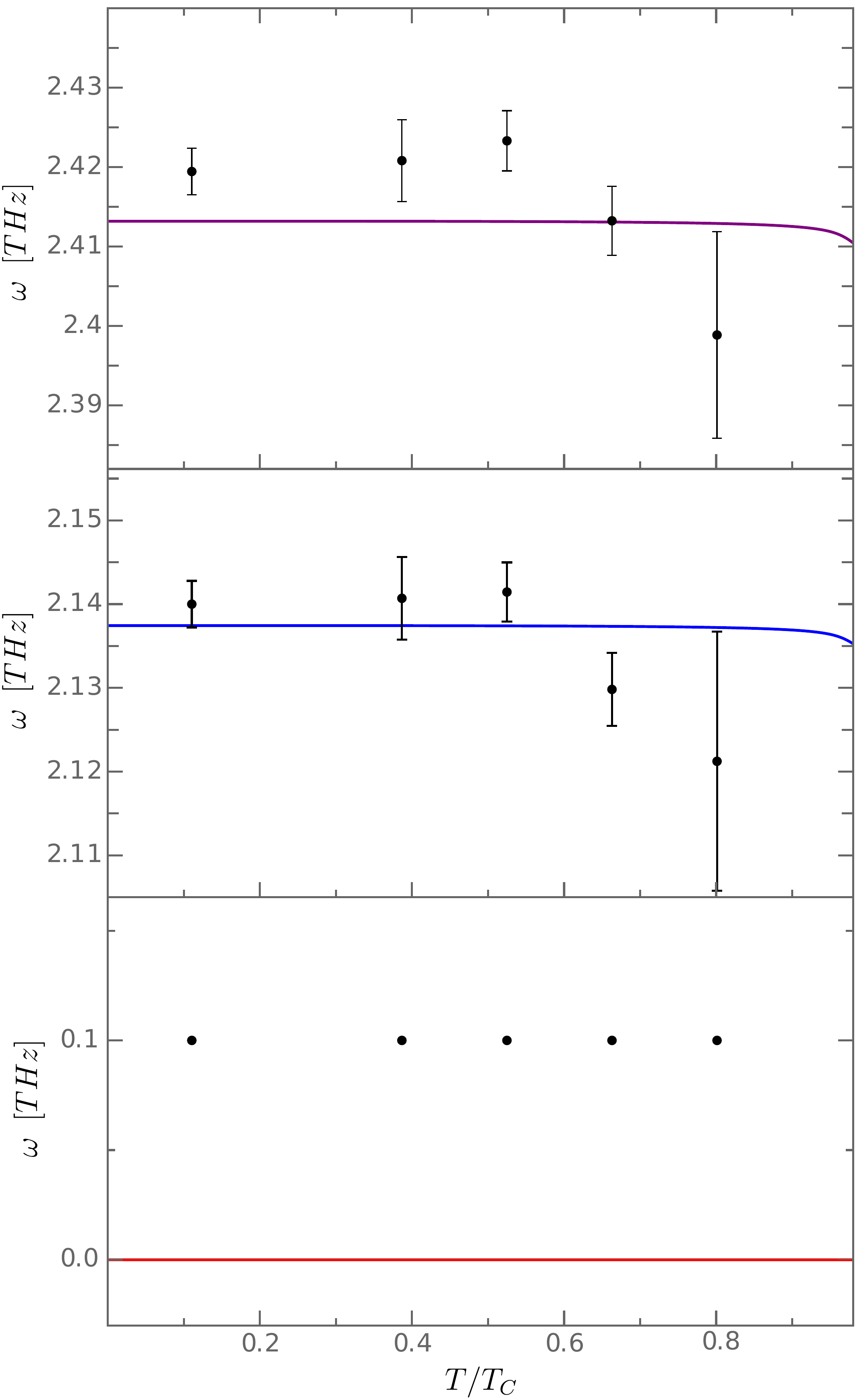}
		\caption{Temperature dependence of the lowest three phase modes of K$_{0.3}$MoO$_3$.  The black dots of the upper two panels are data points of Ref.~\cite{Thomson23} including the experimental uncertainty. The black dots of the lowest panel are taken form the old experiment in \cite{Degiorgi91}, also used for the analyses in \cite{Thomson23}.  The solid lines are the predictions of the mode equation~\eqref{eq:eigenB2}  for $q=0$, 
			using  the fitted parameters $\gamma_{\lambda}^2$ and $\omega_{ 2 \bd{k}_F \lambda}$ 
			from Appendix B. 
		}		\label{fig:ExPhase}
	\end{figure}

	In this work, we retain only the lowest three phase and amplitude modes of K$_{0.3}$MoO$_3$. The frequencies of bare phonon modes for $T > T_c$ are taken from Refs.~\cite{Schaefer10,Schaefer14,Thomson17}, the frequencies of amplitude modes from \cite{Schaefer10,Schaefer14,Thomson17,Thomson23}, and the value of the CDW gap  $2\Delta_0 \approx 100 $ meV is estimated from angle-resolved photoemission measurements \cite{Ando05}, which fixes all the electron-phonon couplings $\gamma_\lambda^2$ for the three relevant phonon modes.  The numerical values of our fit parameters $\omega_{\lambda}$ and $\gamma_{\lambda}^2$ are given in Appendix~B. Then we solve the mean-field gap equation~\eqref{eq:selfcon} to obtain  the temperature-dependent gap $\Delta(T)$. Cross-checking with Refs.~\cite{Sato83,Girault89} reveals that $\Delta(T)$ can be approximated on the mean-field level. Using $\Delta(T)$ as an input for the numerical solution of the function $F(0,\omega )$ defined in Eq.~\eqref{eq:deffnum} enables us to solve for the roots of the mode-frequency equations~\eqref{eq:eigenA2} and \eqref{eq:eigenB2} in the absence of Coulomb interactions for $\Pi^{AA}(\omega+i\delta,0)$ and $\Pi^{BB}(\omega+i\delta,0)$. Our results are presented in Figs.~\ref{fig:ExAmplidude} and~\ref{fig:ExPhase}, where we plot the lowest three temperature-dependent amplitude and phase modes for temperatures up to $T < 0.98 T_c$.	The experimental data were taken in  Refs.~\cite{Schaefer10,Schaefer14,Thomson17,Thomson23,Degiorgi91}. 
	The black dots in Figs.~\ref{fig:ExAmplidude} and~\ref{fig:ExPhase} are a 
	a recently re-analyzed version of the data \cite{Thomson23}.
	
	From Fig.~\ref{fig:ExAmplidude}, we see that 
	our calculations reproduce quantitatively the temperature dependence 
	of the amplitude-mode frequencies from Refs.~\cite{Schaefer10,Schaefer14,Thomson17,Thomson23} as long as $T$ is not too close to $T_c$. 
	When the temperature approaches $T_c$  thermal fluctuations become 
	increasingly important and the order parameter $\Delta(T)$ becomes small so that the Gaussian approximation used in this work is expected to break down. In this regime higher-order processes need to be taken into account, like for example in the theory of fluctuational superconductivity for temperatures close to  $T_c$ \cite{Varlamov_book}. 
	
	In  Fig.~\ref{fig:ExPhase} we compare our theoretical predictions for the energies of the 
	phase modes with the experiments. Although for the finite-frequency modes
	the order of magnitudes agree with \cite{Thomson17,Thomson23}, our 
	Eq.~(\ref{eq:eigenB2}) predicts that the lowest phase mode should be gapless,
	in contrast to the experimental data \cite{Degiorgi91} for the
	lowest phase mode which exhibit  a small but  finite gap. 
	Moreover, our finite-temperature results for the phase modes reveal a discrepancy in the magnitude of the temperature dependence: 
	Our solution of Eq.~\eqref{eq:eigenB2} predicts a  
	variation in the temperature dependence of the finite-frequency phase modes of 
	order $10^{-4}$ THz
	which looks essentially flat on the scale of Fig.~\ref{fig:ExPhase}. 
	By contrast, the experimentally observed  variation, while being generally weak compared to the phase mode frequencies, is still two orders of magnitude larger than the theoretical estimate, i.e., of order $10^{-2}$ THz 
	for the finite frequency phase modes shown in Fig. \ref{fig:ExPhase}. 	A possible explanation for this discrepancy between our theory for the finite frequency phase modes and experiment is that the Gaussian approximation made in deriving Eq.\ \eqref{eq:eigenB2} is not sufficient to explain the temperature dependence of the phase modes. Note that the Gaussian approximation, which is equivalent to the random-phase approximation, retains only the lowest order in the electron-phonon coupling by including only the contribution of a  single electron-bubble to  the phonon self-energy, i.e. it is quadratic in the dimensionless electron-phonon coupling $\sqrt{g_0}$ defined in Eq.~\eqref{eq:gnull}. To this order, there is no mixing between the amplitude and phase modes, see Eq.\ (\ref{eq:SeffAB}). However, the next-order contribution in the collective fields, which is quartic in $\sqrt{g_0}$, already introduces such a mixing that would lead to a stronger temperature variation of the phase modes. Since in K$_{0.3}$MoO$_3$ the dimensionless electron-phonon coupling  $g_0\sim 0.3$ is small, the effect of such mixing can be estimated as $g_0^2 \delta\omega_{A}\sim 0.01$ THz, where we used $\delta \omega_{A} \sim 0.1$ THz  as the variation of the higher frequency amplitude modes in Fig.~\ref{fig:ExAmplidude}. This estimate matches the experimentally observed order of magnitude of the phase mode variation. Hence, a systematic calculation of this effect for the phase modes can be done by considering higher order terms up to the fourth order of the expansion \eqref{eq:expansion1}, posing a route for further improving the accuracy of the theory presented in this work. 
	
	While the experimental data for the amplitude modes and the order of magnitude for the higher-frequency phase modes can be	fitted by our  results for the  multi-phonon Fr\"{o}hlich Hamiltonian for generic electron-phonon interactions, the	behavior of the lowest phase mode remains somewhat mysterious. In fact, in experiments the frequency of this mode has not been unambiguously determined so far; for example, with  the detectors used 	in Ref.~\cite{Thomson17,Thomson23} any mode with frequency of the order of  $1$~THz or lower cannot be reliably detected leading to a difference in the analyses between Ref.~\cite{Thomson17} and Ref.~\cite{Thomson23}.
	Given the fact that according to our analysis the lowest phase mode remains gapless even in the presence of	long-range Coulomb interactions, the physical mechanism inducing a gap in the  	lowest phase mode, seen in an earlier experiment by	Degiorgi {\it{et al.}}~\cite{Degiorgi91}, remains unclear.	One possible explanation is 	pinning by impurities \cite{Kurihara76}, which are not included in our model.	Another possibility is that the gapped lowest-frequency phase mode detected	in the experimental data reproduced in  Fig.~\ref{fig:ExPhase} is actually the upper (gapful) branch of the 	hybrid phason-plasmon mode due to  the Coulomb interaction discussed in Sec.~\ref{sec:Coulomb}.

	\subsection{K$_{0.3}$MoO$_3$ and the Fr\"{o}hlich-Coulomb model}

\begin{figure}
		\centering
		\includegraphics[width=1\linewidth]{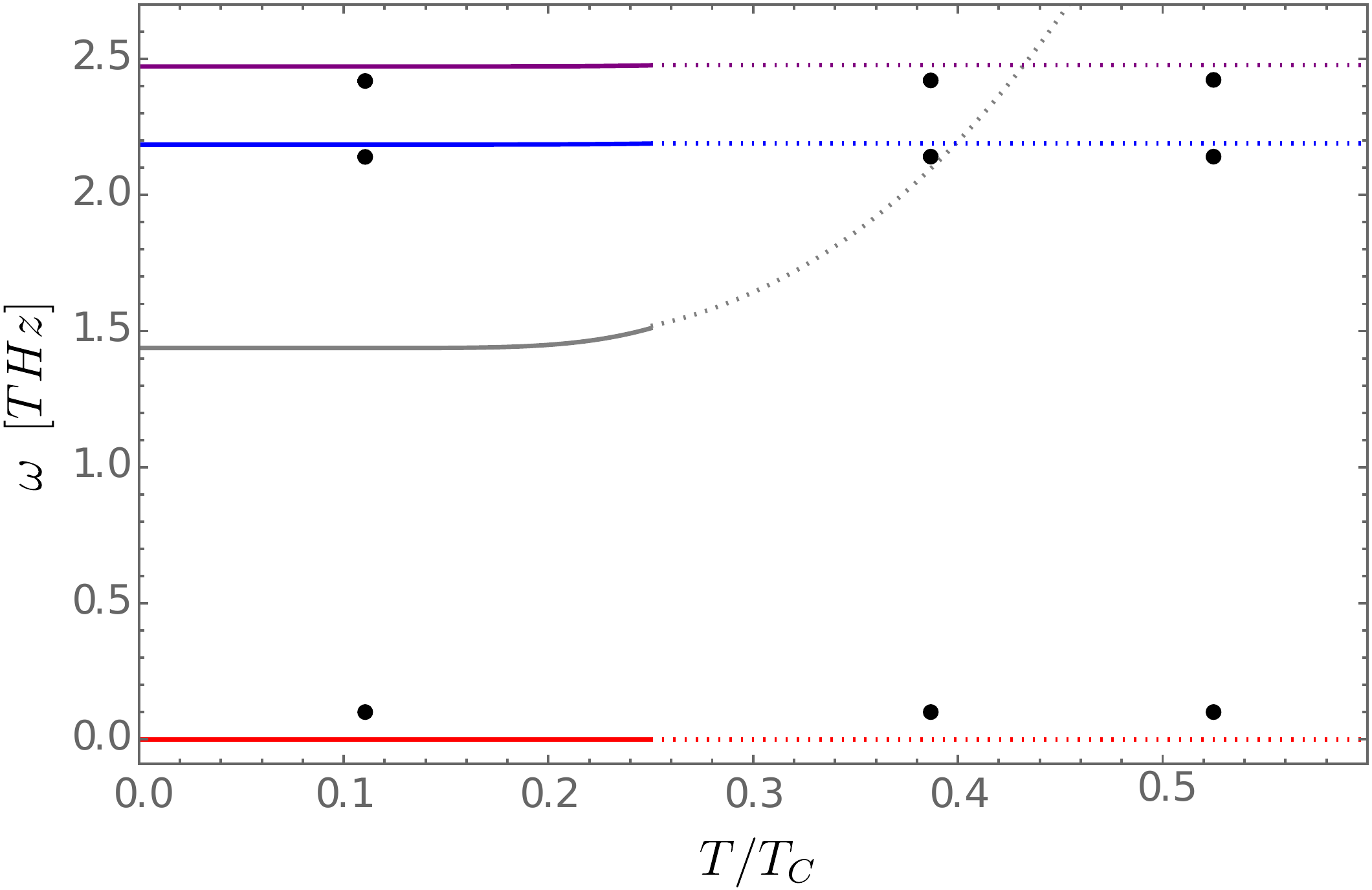}
		\caption{Temperature dependence of the lowest phase modes of K$_{0.3}$MoO$_3$, predicted by the mode equation~\eqref{eq:hybridpp} for $\Omega_e\approx 0.1$ eV, $\Omega_i^2/\Omega_e^2\approx 1/2000$, and $q=0$.  Again the black dots are the experimental data points \cite{Thomson23,Degiorgi91} for the phase modes.  While all modes color graded as in Fig.~\ref{fig:ExPhase} are weakly effected for low temperatures by the inclusion of the additional self-energy correction $\Pi_{1}^{BB}(\omega+i\delta,0)$ due to Coulomb interactions, it gives rise  the hybrid mode $\omega_{+}(0)$ depicted as a gray line here. For $T \gtrsim T_*$ the hybrid mode 
			gains a strong temperature dependence related to the restoration of normal screening at $T=T_c$ as discussed in Appendix B. This behavior, indicated by switching to dotted lines $T \gtrsim T_*$, is not supported by the experimental data~\cite{Thomson23}.}
		\label{fig:PhaseCoulomb}
	\end{figure}

	To examine  the hypothesis that the lowest phase mode is the upper (gapful) branch of the hybrid phason-plasmon mode induced by the Coulomb interaction, we have solved the equation (\ref{eq:hybridpp}) for the collective phase modes including  the effect of  long-range Coulomb interactions. The collective modes now depend on additional parameters, which can be chosen to be the electronic plasma frequency $\Omega_e^2 = 4 \pi  e^2 n /m$ and the  long-wavelength electron-phonon couplings $\alpha_{\bd{q} \lambda} = \alpha_{\lambda} | \bd{q} |$ defined in Eq.~(\ref{eq:alphadef}). For simplicity, we work here with a minimal model where the fluctuations of the ionic density are described by a single acoustic phonon, i.e., neglecting hybridization with the other phonons. Therefore, we can conveniently use  the square of the ionic plasma frequency $f_{\bd{q}} \alpha^2_{\bd{q}}=\Omega_i^2$ as the additional (and experimentally accessible) fit parameter instead of the coupling $\alpha_{\bd{q}}$. With $\Omega_e \approx 0.1$ eV and $\Omega_i^2/\Omega_e^2\approx 1/2000$ we obtain the hybrid phason-plasmon mode indicated by the gray line in Fig.~\ref{fig:PhaseCoulomb}. At the same time, the amplitude modes are not affected by the Coulomb interaction within our approximation and the
	finite frequency phase modes are marginally shifted  by the self-energy correction \eqref{eq:Pi1} due to Coulomb interactions. 
	
	While the frequency $\omega_{+} (0)$ of the hybrid phason-plasmon mode (depicted by the gray line in Fig.~\ref{fig:PhaseCoulomb}) is of a reasonable magnitude and behaves as all the other regular amplitude/phase modes for low enough temperatures, the theoretically predicted  dramatic temperature dependence at higher temperatures disagrees with the experiments \cite{Thomson17,Thomson23}.  For $T$ exceeding a certain temperature $T_{\ast}$ (which we estimate in Appendix~B) the temperature dependence of our theoretical expression for $\omega_{+} (0)$, shown  in Fig.~\ref{fig:PhaseCoulomb}, is due to a strongly $T$-dependent contribution 	to  the polarization function $\Pi^{\phi\phi}_0(Q)$ defined in Eq.~\eqref{eq:Piphiphiq2}. 	This contribution, which vanishes  for $T\rightarrow 0$, restores  to the usual Lindhard function \cite{Fetter71,Mihaila11} in the normal metallic state and dominates the temperature dependence of the hybrid mode for $T \gtrsim T_{\ast}$. This contribution 	was already taken into  account in the calculations including long-range Coulomb interactions by Visosztek and Maki \cite{Virosztek93}, yet because they neglect the ionic response it plays a different role there.

	One possible explanation for the above discrepancy between theory and experiment is the breakdown of the Gaussian approximation (which is equivalent to the random-phase approximation)  to properly describe the formation of charge-density waves  in the presence of  Coulomb interactions for $T \gtrsim T_*$. 
	Since  Luttinger liquid physics is generally expected to become important in electronically 
	one-dimensional systems with Coulomb interaction~\cite{Giamarchi_Book}, higher-order interaction processes neglected in Gaussian approximation are expected to stabilize the CDW state~\cite{Anderson70,Dzyaloshinskii74,Solyom79} in the temperature range $T \gtrsim T_*$, 
	and suppress the Fermi liquid physics  including metallic screening described by the temperature dependence of polarization function $\Pi^{\phi\phi}_0(Q)$ for $T \gtrsim T_*$. We therefore  suggest that the proper  theory of collective modes in the presence of long-range Coulomb interactions for $T>T_*$ would need to include interactions between the collective modes, which is a possible further theoretical development but goes beyond the scope of the present work.

	\section{Summary and conclusions}
	
	In this work, we have re-examined the effect of long-range Coulomb interactions on the collective amplitude and phase modes in an electronically one-dimensional incommensurate CDW. We have shown that the lowest phase mode  has a gapless linear dispersion even in the presence of long-range Coulomb interactions. Our calculation reveals the crucial role of the longitudinal acoustic phonon to protect the gapless nature of the lowest phase mode in the presence of long-range Coulomb interactions. This is an interesting example of the interplay of two Goldstone modes associated with the spontaneous breaking of two different continuous symmetries in a solid. Previous investigations of this problem \cite{Wong87,Virosztek93} have not properly taken the fluctuations of the ionic charge density and the associated contribution to the dielectric function into account. Let us emphasize that a correct description of screening is crucial to calculate the collective modes in a CDW: Although the electronic contribution to the dielectric function is finite due to the CDW gap, the ionic contribution associated with the last term in our expression (\ref{eq:dielectricfunc}) for the dielectric function restores the usual $1/q^2$ divergence of the static dielectric function due to the contribution from the acoustic phonon. As a result, the Coulomb interaction is effectively screened and the phase mode remains gapless. If we incorrectly omit the contribution of the acoustic phonon in the sum $\sum_{\lambda}  \alpha_{\bd{q} \lambda}^2 /( \nu^2 + \omega_{\bd{q} \lambda}^2 )$ in Eq.~(\ref{eq:dielectricfunc}) this contribution would vanish as $ \bd{q}^2$ for small $\bd{q}$ leading to a finite static dielectric function; the Coulomb interaction would then remain long-range leading to a gapped phase mode. 
	
	Comparison with experiments  on K$_{0.3}$MoO$_3$ (blue bronze) \cite{Thomson17,Schaefer14,Schaefer10,Thomson23} shows that the temperature-dependent mode frequencies of multiple amplitude modes are already well described by the multi-phonon Fr\"{o}hlich Hamiltonian for generic electron-phonon interactions in Gaussian approximation derived in Sec.~\ref{sec:modesF}. At this level of approximation, our effective action approach is essentially equivalent to the random-phase approximation adopted long time ago by Lee, Rice, and Anderson~\cite{Lee73,Lee74,Rice75,Rice76,Rice78}. While this  model is fairly successful in describing the order of magnitude for the higher frequency phase modes, our calculations for the multi-phonon Fr\"{o}hlich model for generic electron-phonon interactions (\ref{eq:Froehlich}) and its generalization (\ref{eq:FroehlichC}) including the long-range Coulomb interaction fail to ascertain the nature of the lowest phase mode detected in Ref.~[\onlinecite{Thomson17}, \onlinecite{Thomson23,Degiorgi91}] and exhibit a temperature dependence far beneath the experimental values. At this point, we can only speculate about the nature of the experimentally observed lowest finite frequency phase  mode and if  a stabilized version of the $T=0$ phason-plasmon hybrid mode predicted by Eq.~\eqref{eq:SeffABCB} exists. We note however, that pinning by impurities would raise the zero-frequency phase mode to finite frequencies. \\
	Finally, let us point out that the experimental data \cite{Schaefer10,Schaefer14,Thomson17,Thomson23} for the amplitude modes and the phase modes exhibit a finite broadening. Unfortunately, at the level of the Gaussian approximation, the collective modes predicted by  Eq.~\eqref{eq:SeffAB} or Eq.~\eqref{eq:SeffABCB} have infinite lifetime and hence do not exhibit any broadening. To compute the damping of the collective modes  within our effective action approach one has to go beyond the Gaussian approximation, retaining in an expansion in powers of the electron-phonon coupling $g_0$ terms up to order $g_0^3$ for the amplitude-  and up to order  $g_0^4$ for the phase modes \cite{Kurihara80,Takada85}. An alternative method of calculating the damping of the collective modes based on the solution of kinetic equations will be presented in Ref.~[\onlinecite{Hahn23}].

	\section*{ACKNOWLEDGEMENTS}	
	
	We are grateful to the Deutsche Forschungsgemeinschaft (DFG, German Research Foundation) for financial support via  TRR 288 - 422213477 (Projects A07, B08 and B09) and via Project No. 461313466.

	\appendix
	
	\setcounter{equation}{0}
	
	\renewcommand{\theequation}{A\arabic{equation}}
	
	\renewcommand{\appendixname}{APPENDIX}
	
	\renewcommand{\thesection}{\Alph{section}}

	\section*{Appendix A: Effective action with Coulomb interactions}
	
	\label{app:static}
	
	\renewcommand{\theequation}{A\arabic{equation}}
	
	In this appendix we outline the manipulations leading from
	the combined action  $S_{\rm eff} [ X , \phi ]$
	of the phonons and the Coulomb field given in Eq.~(\ref{eq:SeffXphi})
	to the effective action $S_{\rm eff} [ A , B ]$ of the collective amplitude and phase fluctuations
	in the presence of long-range Coulomb interactions given in Eq.~(\ref{eq:SeffABCB}).
	Starting from Eq.~(\ref{eq:SeffXphi}) 
	we expand the last term up to quadratic order 
	in the fluctuations and obtain the Gaussian low-energy effective action
	\begin{eqnarray}
		S_{\rm eff} [ U, V , \phi ] & = &   \beta \Omega_1 + \beta {\cal{V}} \sum_{\lambda} \gamma_{\lambda} | X^0_{\lambda} |^2
		\nonumber
		\\
		& + & 
		\frac{1}{2} \int_{ Q \lambda} ( \nu^2 + \omega_{\bd{q} \lambda}^2 ) U_{ - Q \lambda } U_{ Q \lambda} 
		\nonumber
		\\
		& + & 
		\int_{ Q \lambda} ( \nu^2 + \omega_{\lambda}^2 ) V^{\ast}_{  Q \lambda } V_{ Q \lambda} 
		\nonumber
		\\
		& + &  \frac{1}{2} \int_Q f_{\bd{q}}^{-1} \phi_{ - Q } \phi_Q + i \int_Q \phi_{ - Q } U_{Q}
		\nonumber
		\\
		& + & \frac{s}{2}  {\rm Tr}( \mathbf{G}_1 \mathbf{V}   \mathbf{G}_1 \mathbf{V}    ),
	\end{eqnarray}
	where we have used the same notation as in Sec.~\ref{sec:modesF}.
	The trace in the last term can be written as
	\begin{eqnarray}
		& & \frac{s}{2} {\rm Tr} [ \mathbf{G}_{1} \mathbf{V}  \mathbf{G}_{1} \mathbf{V}  ] 
		\nonumber
		\\ 
		&
		= &  \int_Q \Bigl[ \frac{1}{2} \Pi_0^{ \phi \phi } ( Q ) \phi_{-Q} \phi_Q   +    \Pi_0^{ \bar{V} V } ( Q ) V^{\ast}_Q V_Q 
		\nonumber
		\\
		& & 
		+ \frac{ \Pi_0^{VV} ( Q ) }{2} V_{-Q } V_Q
		+ \frac{ \Pi_0^{\bar{V} \bar{V}} ( Q ) }{2} V^{\ast}_{Q } V^{\ast}_{-Q}		\nonumber
		\\
		& & + i \Pi_0^{ \phi V} ( Q ) \phi_{ - Q } V_Q + i \Pi_0^{ \phi \bar{V}} ( Q ) \phi_{ -Q} V^{\ast}_{ - Q } 
		\Bigr],
	\end{eqnarray}
	where the generalized polarization functions $\Pi_0^{\bar{V} V} (Q)$, $\Pi_0^{ VV} (Q)$, and 
	$\Pi_0^{ \bar{V} \bar{V}} (Q)$ are defined in Eq.~(\ref{eq:pidefs}) and we have introduced three more polarization functions associated with the Coulomb field $\phi$,
	\begin{widetext}
		\begin{subequations}
			\label{eq:pidefs2}
			\begin{eqnarray}
				\Pi_0^{\phi \phi} ( \bd{q} , i \nu  ) & = & -2 s \int_K
				\frac{ i \omega ( i \omega + i \nu ) +  v_F^2 k_x ( k_x +q_x ) + | \Delta |^2 }{
					[ ( i \omega )^2 - ( v_F k_x )^2 - | \Delta |^2 ]
					[ ( i \omega + i \nu )^2 - ( v_F (k_x + q_x))^2 - | \Delta |^2 ] },
				\\
				\Pi_0^{ \phi {V}} ( \bd{q} , i \nu  ) & = & s
				\int_K \frac{ v_F q_x \Delta^{\ast} }{
					[ ( i \omega )^2 - ( v_F k_x )^2 - | \Delta |^2 ]
					[ ( i \omega + i \nu )^2 - ( v_F (k_x + q_x))^2 - | \Delta |^2 ] },
				\\
				\Pi_0^{ \phi \bar{V}} ( \bd{q} , i \nu  ) & = & - s
				\int_K \frac{ v_F q_x \Delta }{
					[ ( i \omega )^2 - ( v_F k_x )^2 - | \Delta |^2 ]
					[ ( i \omega + i \nu )^2 - ( v_F (k_x + q_x))^2 - | \Delta |^2 ] }.
			\end{eqnarray}
		\end{subequations}
		Since we are only interested in the phonons, we now
		integrate over the Coulomb field. The resulting effective phonon action is
		\begin{eqnarray}
			S_{\rm eff} [ U, V ]  &= &  \beta \Omega_1 + \beta {\cal{V}} \sum_{\lambda} \gamma_{\lambda} | X^0_{\lambda} |^2
			+
			\frac{1}{2} \int_{ Q \lambda} ( \nu^2 + \omega_{\bd{q} \lambda}^2 ) U_{ - Q \lambda } U_{ Q \lambda} 
			\nonumber
			\\
			& + & 
			\int_{ Q } \biggl[  \sum_{\lambda}
			( \nu^2 + \omega_{\lambda}^2  ) V^{\ast}_{  Q \lambda } V_{ Q \lambda}  
			+ \Pi_0^{ \bar{V} V} ( Q ) V^{\ast}_Q V_Q 
			+ \frac{ \Pi_0^{VV} ( Q ) }{2} V_{-Q } V_Q
			+ \frac{ \Pi_0^{\bar{V} \bar{V}} ( Q ) }{2} V^{\ast}_{Q } V^{\ast}_{-Q} \biggr] 
			\nonumber
			\\
			& + & \frac{1}{2} \int_Q \frac{ f_{\bd{q}}}{ 1 + f_{\bd{q}} \Pi_0^{\phi \phi} ( Q ) }
			[ U_{ - Q } + \Pi_0^{\phi V} ( - Q ) V_{ - Q } + \Pi_0^{ \phi \bar{V}} ( - Q )  V^{\ast}_Q ]
			[ U_{  Q } + \Pi_0^{\phi V} (  Q ) V_{  Q } + \Pi_0^{ \phi \bar{V}} (  Q )  V^{\ast}_{-Q} ].
			\label{eq:SeffUVcb}
		\end{eqnarray}
		At this point we use the gauge freedom choose $\Delta$ to be real and 
		express the complex fields $V_{Q \lambda}$ in terms of two real fields as
		in Eq.~(\ref{eq:ABdef}). Then Eq.~(\ref{eq:SeffUVcb}) can be written as		\begin{eqnarray}
			S_{\rm eff} [ U_{\lambda} , A_{\lambda} , B_{\lambda} ]  &= &  \beta \Omega_1 + \beta {\cal{V}} \sum_{\lambda} \gamma_{\lambda} | X^0_{\lambda} |^2
			+
			\frac{1}{2} \int_{ Q } \sum_{\lambda} ( \nu^2 + \omega_{\bd{q} \lambda}^2 ) U_{ - Q \lambda } U_{ Q \lambda} 
			\nonumber
			\\
			& + & 
			\frac{1}{2} \int_{ Q } \biggl[  \sum_{\lambda} ( \nu^2 + \omega_{\lambda}^2 ) ( A_{-Q \lambda} A_{Q  \lambda} + B_{ - Q \lambda} B_{ Q \lambda} ) + \Pi_0^{ AA} ( Q) A_{ -Q} A_Q + \Pi_0^{ BB} ( Q) B_{ -Q} B_Q
			\biggr]
			\nonumber
			\\
			& + & \frac{1}{2} \int_Q \frac{ f_{\bd{q}}}{ 1 + f_{\bd{q}} \Pi_0^{\phi \phi} ( Q ) }
			[ U_{ - Q } + \sqrt{2} i \Pi_0^{\phi V} ( - Q ) B_{ - Q }  ]
			[ U_{  Q } + \sqrt{2} i \Pi_0^{\phi V} (  Q ) B_{  Q } ].
		\end{eqnarray}
		Note that the Coulomb interaction couples the phase mode $B$ to the long-wavelength phonon $U$.
		The energies of the collective modes can be obtained from the 
		effective action for collective fields $U_Q = \sum_{\lambda} \alpha_{\bd{q} \lambda} U_{Q \lambda}$,
		$A_{ Q } = \sum_{\lambda} \gamma_{\lambda} A_{Q \lambda}$ and
		$B_{ Q } = \sum_{\lambda} \gamma_{\lambda} B_{Q \lambda}$. 
		As in Eq.~(\ref{eq:HSaverage}) we implement 
		the constraints via auxiliary fields and obtain
		\begin{eqnarray}
			S_{\rm eff} [ U , A , B ]  &= &  \beta \Omega_1 + \beta {\cal{V}} \sum_{\lambda} \gamma_{\lambda} | X^0_{\lambda} |^2
			+
			\frac{1}{2} \int_{ Q } \bar{D}_{0}^{-1} ( Q)  U_{ - Q  } U_{ Q } 
			\nonumber
			\\
			& + & 
			\frac{1}{2} \int_{ Q } \biggl[   \left[  
			\bar{D}_0^{-1} ( 2 k_F , i \nu ) + \Pi_0^{ AA} ( Q) \right] A_{ -Q} A_Q     
			+ \left[   \bar{D}_0^{-1} ( 2 k_F , i \nu )   + \Pi_0^{ BB} ( Q) \right] B_{ -Q} B_Q
			\biggr]
			\nonumber
			\\
			& + & \frac{1}{2} \int_Q \frac{ f_{\bd{q}}}{ 1 + f_{\bd{q}} \Pi_0^{\phi \phi} ( Q ) }
			[ U_{ - Q } + \sqrt{2} i \Pi_0^{\phi V} ( - Q ) B_{ - Q }  ]
			[ U_{  Q } + \sqrt{2} i \Pi_0^{\phi V} (  Q ) B_{  Q } ],
		\end{eqnarray}
	\end{widetext}
	where
	$\bar{D}_0 ( Q ) = \sum_{\lambda} \frac{ \alpha^2_{ \bd{q} \lambda}}{ \nu^2 + \omega^2_{\bd{q} \lambda} }$ and  $\bar{D}_0 ( 2 k_F , i \nu )= \sum_{\lambda} \frac{ \gamma^2_{\lambda}}{ \nu^2+ \omega_{\lambda}^2 }$.	Since we are only interested in the collective amplitude and phase modes, 
	we may now integrate over the
	fluctuations $U_Q$ of the long-wavelength phonons. Dropping  field-independent
	constants we obtain the effective action $S_{\rm eff} [ A , B ]$
	given in Eq.~(\ref{eq:SeffABCB}) of the main text.
	
	\section*{Appendix B: Fit parameters}
	
	\setcounter{equation}{0}
	\renewcommand{\theequation}{B\arabic{equation}}

	In this appendix  we specify the parameters in the multi-phonon Fr\"{o}hlich Hamiltonian for generic electron- phonon interactions(\ref{eq:Froehlich}) and in the Fr\"{o}hlich-Coulomb Hamiltonian (\ref{eq:FroehlichC}) which we have used in Sec.~\ref{sec:experiment} to compare our calculations  with  the experimental data. The multi-phonon Fr\"{o}hlich Hamiltonian for generic electron-phonon interactions~\eqref{eq:Froehlich} with $\lambda_{\rm max}$ phonons depends on the bare phonon frequencies $\omega_{ 2 \bd{k}_F \lambda}$ and  the electron-phonon couplings $\gamma_{\lambda}$, which form a set of $2 \lambda_{\rm max}$ parameters. In addition, out Hamiltonian depends also on the ultraviolet  cutoff $\omega_c$ in the mean-field gap equation~(\ref{eq:gapequation}); actually, we may use the gap equation (\ref{eq:gapequation}) to eliminate $\omega_c$ in favor of the experimentally measured zero-temperature gap $2\Delta (T=0) \approx 100$~meV$ \approx  24$ THz. It is then natural to define the dimensionless couplings
	\begin{equation}
		\bar{\gamma}_{\lambda}^2
		=\frac{\gamma_{\lambda}^2\nu_3}{4 \Delta^2(T=0)}, \; \; \; \; \; \; 		g_{0,\lambda}=\frac{\gamma_{\lambda}^2\nu_3}{2\omega^2_{ 2 \bd{k}_F  \lambda}}.
		\label{eq:couplingsdimless}
	\end{equation} 
	Taking the values for $\omega_{ 2 \bd{k}_F  \lambda}$ from the supplementary material of Ref.~[\onlinecite{Thomson17}] 
	for $T > T_c$ and the amplitude modes from \cite{Schaefer10,Schaefer14,Thomson23} 
	we tune the $\bar{\gamma}_{\lambda}^2$ to match the solution of the  amplitude mode equation~\eqref{eq:eigenA2} to its measured counterparts at $T = 0$. A summary of those values are shown in Table~\ref{fig:werte}. Retaining in the Fr\"{o}hlich-Coulomb Hamiltonian \eqref{eq:FroehlichC} only the  longitudinal acoustic phonon, we obtain  for small momenta
	\begin{equation}\label{eq:appendixD0}
		f_{\bd{q}}\bar{D}_0(Q) \approx 
		\frac{ f_{\bd{q}}\alpha^2_{ \bd{q} }}{ \nu^2 + \omega^2_{\bd{q}} }=\frac{\Omega_i^2}{\nu^2 + \omega^2_{\bd{q}}},
	\end{equation}
	with $\omega_{\bd{q}} =  v_s |\bd{q}|$. Therefore, the minimal Fr\"{o}hlich-Coulomb Hamiltonian depends on  two additional parameters which can be taken to be the ionic and electronic plasma frequencies,  $\Omega_i$ and $\Omega_e$, 
	specified in the caption of Fig.~\ref{fig:PhaseCoulomb}.
	\vspace{2mm}
	\begin{table}[h!]
		\begin{center}
			\begin{tabular}{|c||c|c|c|}
				\hline
				\hline
				mode number & 1 & 2 & 3 \\
				\hline
				\hline
				$\omega_{ 2 \bd{k}_F  \lambda}$ [THz] & 1.79 & 2.25 & 2.64 \\
				\hline
				$\omega_{A,\lambda}(T=0)$ [THz] & 1.69 & 2.23 &	2.56\\
				\hline
				$\omega_{B,\lambda} (T = 0)$ [THz] & 0.10 &	2.14 & 2.41 \\
				\hline
				$\bar{\gamma}^2_{\lambda}\times10^3$ (fitted) & 2.30 & 0.75 & 3.74 \\
				\hline
				$g_{0,\lambda}$ (fitted) & 0.42 & 0.09 & 0.31 \\
				\hline\hline
			\end{tabular}
		\end{center}
		\caption{Summary of the values of the parameters of  the multi-phonon Fr\"{o}hlich Hamiltonian for generic electron-phonon interactions (\ref{eq:Froehlich}) and the 
			dimensionless coupling constants defined in Eq.~(\ref{eq:couplingsdimless}) used for the fit to the experiments.}
		\label{fig:werte}
	\end{table}

	Finally, let us estimate the temperature scale $T_{\ast}$ above which, according to the discussion at the end of  Sec.~\ref{sec:experiment}, non-Gaussian corrections to the polarization functions become important. Therefore we calculate the polarization function $\Pi_0^{\phi \phi} ( \bd{q} , i \nu )$ defined in Eq.~\eqref{eq:pidefs2}  for $T > 0$. After carrying out the frequency sum we obtain
	\begin{widetext}
		\begin{align}
			\Pi_0^{\phi\phi}( \bd{q} ,i\nu)=-\frac{s}{\cal{V}}& \sum_{\bd{k}} 
			\left[1+\frac{v_F^2k_x(k_x+q_x)+|\Delta|^2}{E_{\bd{k}} E_{\bd{k} + \bd{q}}}\right]
			\frac{(f_{\bd{k} +\bd{q}}-f_{\bd{k}})(E_{ \bd{k}+ \bd{q}}-E_{\bd{k}})}{(E_{\bd{k}+ \bd{q}}-E_{\bd{k}})^2+\nu^2}
			\nonumber
			\\
			+\frac{s}{\cal{V}} &\sum_{\bd{k}}\left[1-\frac{v_F^2k_x(k_x+q_x)+|\Delta|^2}{E_{\bd{k}} 
				E_{ \bd{k}+ \bd{q}}}\right]\frac{(1-f_{ \bd{k}+\bd{q}}-f_{\bd{k}} )(E_{ \bd{k}+\bd{q}}+E_{\bd{k}})}{
				(E_{ \bd{k}+ \bd{q}}+E_{ \bd{k}})^2+\nu^2},
		\end{align}
	\end{widetext}	
	where for an electronically one-dimensional system $E_{\bd{k}} = \sqrt{ ( v_F k_x )^2 + \Delta^2 }$ and
	$f_{\bd{k}} = 1/( e^{\beta E_{\bd{k}}} +1 )$ is the Fermi function. In the regime $v_F | q_x |  \ll \omega$ this 
	becomes
	\begin{align}
		\Pi^{\phi\phi}_0(q_x,i\nu)\approx - 2s \frac{(v_F q_x )^2}{\nu^2 {\cal{V}} }\sum_{\bd{k}} \frac{(v_F k_x)^2}{E_{\bd{k}}^2}f'_{\bd{k} } \nonumber  \\
		+\frac{s}{4} \frac{ (v_F q_x)^2}{\cal{V}} \sum_{\bd{k}}\frac{ \Delta^2  }{E_{\bd{k}}^5}\tanh\left(\frac{\beta 
			E_{\bd{k}}}{2}\right).
		\label{eq:Piphiphiq2}
	\end{align}
	
	\noindent At zero temperature Eq.~\eqref{eq:Piphiphiq2} collapses to Eq.~\eqref{eq:PiphiphiT0}.  On the other hand, at the critical temperature where $\Delta ( T_c) =0$ the first term involving the derivative of the Fermi function reduces to the usual high-frequency behavior of the Lindhard function in the limit $v_Fq\ll\omega$, whereas the second term vanishes. The crossover scale $T_{\ast}$ can be estimated from the condition that the first term on the right-hand side of Eq.~(\ref{eq:Piphiphiq2}) has the same order of magnitude as the sum of the second term and the ionic screening Eq.~\eqref{eq:appendixD0}. In terms of the dimensionless function 
	\begin{equation}
		A(T)=- \frac{2\pi v_F}{\cal{V}} \sum_{\bd{k}}\frac{(v_F k_x)^2}{E_{\bd{k}}^2}f'_{\bd{k}}
	\end{equation}
	this condition can be written as $A ( T_{\ast} ) /\omega^2_+ ( 0 ) = 1/ \Delta^2 ( T_{\ast} )$.
	Using our result (\ref{eq:omegaplusdelta}) this leads to the condition	
	\begin{equation}\label{eq:P1exp}
		\frac{\Omega_i^2}{\Omega_e^2}+\frac{v_\varphi^2}{v_F^2}\approx A(T_\ast),
	\end{equation}
	which defines the non-universal crossover temperature $T_{\ast}$ above which the Gaussian approximation 
	used in this work is likely to break down  in the presence of Coulomb interactions.	
	Substituting the experimentally relevant parameters for
	K$_{0.3}$MoO$_3$ we estimate $T_{\ast} \approx 0.3 T_c$, as indicated by the upturn of the
	dotted line in Fig.~\ref{fig:PhaseCoulomb}.


\begin{thebibliography}{99}
		%
		\bibitem{Lee73}
		P. A. Lee, T. M. Rice, and P. W. Anderson,
		Fluctuation Effects at a Peierls Transition,
		Phys. Rev. Lett. {\bf{31}}, 462 (1973).
		%
		\bibitem{Lee74}
		P. A. Lee, T. M. Rice, and P. W. Anderson, 
		Conductivity from Charge or Spin Density Waves,
		Solid State Commun. {\bf{14}}, 703 (1974).
		\bibitem{Rice75}
		M. J. Rice, C. B. Duke, and N. O. Lipari,
		Intermolecular vibrational stabilization of the charge density wave state in organic metals,
		Solid State Commun. {\bf{17}}, 1089 (1975).
		%
		\bibitem{Rice76}
		M. J. Rice, Organic Linear Conductors as Systems for the Study
		of Electron-Phonon Interactions in the Organic Solid State,
		Phys. Rev. Lett. {\bf{37}}, 36 (1976).
		%
		\bibitem{Rice78}
		M. J. Rice, Dynamical Properties of the Peierls-Fr\"{o}hlich State on the Many-Phonon-Coupling Model,
		Solid State Commun. {\bf{25}}, 1083 (1978).
		%
		\bibitem{Gruener88}
		G. Gr\"{u}ner, The dynamics of charge-density waves,
		Rev. Mod. Phys. {\bf{60}}, 1129 (1988).
		%
		\bibitem{Gruener94}
		G. Gr\"{u}ner, {\it{Density waves in solids}}, (Addison-Wesley Frontiers in
		Physics, 
		Reading, MA, 1994).
		%
		\bibitem{Pouget91}
		J.-P. Pouget, B. Hennion, C. Escribe-Filippini, and M. Sato, Neutron Scattering Investigations of the Kohn Anomaly and of the Phase and Amplitude Charge Density Wave Excitations of the Blue Bronze K$_{0.3}$MoO$_{3}$, Phys. Rev. B. {\bf{43}}, 8421 (1991).
		%
		\bibitem{Hennion92}
		B. Hennion, J.-P. Pouget, and M. Sato, Charge-Density-Wave Phase Elasticity of the Blue Bronze, Phys. Rev. Lett. {\bf{68}}, 2374 (1992).
		%
		\bibitem{Ravy04}
		S. Ravy, H. Requardt, D. Le Bolloc'h, P. Foury-Leylekian, J.-P. Pouget, R. Currat, P. Monceau, and M. Krisch, Inelastic X-ray scattering study of CDW dynamics in the Rb0.3MoO3 blue bronze, Phys. Rev. B 69, 115113 (2004).
		%
		\bibitem{Schaefer13}
		H. Schaefer , M. Koerber , A. Tomeljak , K. Biljakovic, H. Berger, and J. Demsar, Dynamics of charge density wave order in the quasi one dimensional conductor (TaSe$_{4}$)$_{2}$I probed by femtosecond optical spectroscopy, Eur. Phys. J. Spec. Top. {\bf 222}, 1005 (2013).
		%
		\bibitem{Kim22}
		S. Kim, Y. Lv, X.-Q. Sun, C. Zhao, N. Bielinski, A. Murzabekova, K. Qu, R. A. Duncan, Q. L. D. Nguyen,
		M. Trigo, D. P. Shoemaker, B. Bradlyn, and F. Mahmood, Observation of a massive phason in a
		charge-density-wave insulator, Nat. Mater. {\bf 22}, 429 (2023).
		%
		\bibitem{Nguyen22}
		Q. L. Nguyen, R. A. Duncan, G, Orenstein, Y. Huang, V. Krapivin, G. de la Pena, C. Ornelas-Skarin, D. A. Reis, P. Abbamonte, S, Bettler, M. Chollet, M. C. Hoffmann, M. Hurley, S. Kim, P. S. Kirchmann, Y. Kubota, F. Mahmood, A. Miller, T. Osaka, K. Qu, T. Sato, D. P. Shoemaker, N. Sirica, S. Song, J. Stanton, S. W. Teitelbaum, S. E. Tilton, T. Togashi, D. Zhu, and M. Trigo, Ultrafast x-ray scattering reveals composite amplitude collective mode in the Weyl charge density wave material (TaSe$_{4}$)$_{2}$I,
		arXiv:2210.17483v2 [cond-mat.mtrl-sci] 23 Dec 2022.
		%
		\bibitem{Schaefer10}
		H. Schaefer, V. V. Kabanov, M. Beyer, K. Biljakovic, and J. Demsar,
		Disentanglement of the Electronic and Lattice Parts of the Order Parameter in a      1d Charge Density Wave System Probed by Femtosecond Spectroscopy,
		Phys. Rev. Lett. {\bf 105}, 066402 (2010).
		%
		\bibitem{Schaefer14}
		H. Schaefer, V. V. Kabanov, and J. Demsar,
		Collective modes in quasi-one-dimensional charge-density wave systems probed by      femtosecond time-resolved optical studies,
		Phys. Rev. B {\bf 89}, 045106 (2014).
		
		\bibitem{Thomson17}
		M. D. Thomson, K. Rabia, F. Meng, M. Bykov, S. van Smaalen, and H. G.
		Roskos, Phase-channel dynamics reveal the role of impurities and screening in a      quasi-one-dimensional charge-density wave system,
		Sci. Rep. {\bf{7}}, 2039 (2017).
		%
		\bibitem{Thomson23}
		K. Warawa, N. Christophel, S. Sobolev, J. Demsar, H. G. Roskos, and M. D. Thomson,
		Combined investigation of collective amplitude and phase modes in a quasi-one-dimensional charge-density-wave system over a wide spectral range,
		submitted to Phys. Rev. B (March 2023).
		%
		\bibitem{Lee78}
		P. A. Lee and H. Fukuyama,
		Dynamics of charge-density wave. II. Long-range Coulomb effects in an array of 
		chains, Phys. Rev. B {\bf{17}}, 542 (1978).
		%
		\bibitem{Wong87}
		K. Y. M. Wong and S. Takada,
		Effects of quasiparticle screening on collective modes: Incommensurate
		charge-density-wave systems, 
		Phys. Rev. B {\bf{36}}, 5476 (1987).
		%
		\bibitem{Virosztek93}
		A. Visosztek and K. Maki,
		Collective modes in charge-density waves and long-range Coulomb interactions,
		Phys. Rev. B {\bf{48}}, 1368 (1993).
		%
		\bibitem{Froehlich52}
		H. Fr\"{o}hlich, Interaction of electrons with lattice vibrations,
		Proc. Roy. Soc. A {\bf{215}}, 291 (1952).
		%
		\bibitem{Veuillen87}
		J. Y. Veuillen, R. C. Cinti and E. Al Khoury Nemeh,
		Direct Determination of Fermi Wave Vector in the Blue Bronze by Means of Angle-Resolved   
		photoemission Spectroscopy,
		Europhys. Lett. {\bf{3}}, 355 (1987).
		%
		\bibitem{Brovman67}
		E. G. Brovman and Yu. Kagan, The phonon spectrum of metals,  Sov. Phys. JETP {\bf{25}},
		365 (1967).
		%
		\bibitem{Gelikman75}
		B. T. Gelikman, Adiabatic perturbation theory for metals and the
		problem of lattice instability, Sov. Phys.  Usp. {\bf{18}},  190
		(1975).
		%
		\bibitem{Yuzbashyan22}
		E. A. Yuzbashian and B. L. Altshuler,
		Breakdown of the Migdal-Eliashberg Theory and a Theory of Lattice-Fermionic Superfluidity, Phys. Rev B {\bf{106}}, 054518 (2022).
		%
		\bibitem{Kumar08}
		S. Kumar and J. van den Brink, 
		Charge ordering and magnetism in quarter-filled Hubbard-Holstein model,		Phys. Rev. B {\bf{78}}, 155123 (2008).
		%
		\bibitem{Murakami14}
		Y. Murakami, P. Werner, N. Tsuji, and H. Aoki,
		Supersolid Phase Accompanied by a Quantum Critical Point in the Intermediate Coupling
		Regime of the Holstein Model, Phys. Rev. Lett. {\bf{113}}, 266404 (2014).
		%
		\bibitem{Ohgoe17}
		T. Ohgoe and M. Imada, Competition among Superconducting, Antiferromagnetic, and Charge Orders with
		Intervention by Phase Separation in the 2D Holstein-Hubbard Model,
		Phys. Rev. Lett. {\bf{119}}, 197001 (2017).
		%
		\bibitem{Esterlis18}
		I. Esterlis, B. Nosarzewski, E. W. Huang, B. Moritz, T. P. Devereaux, D. J. Scalapino, and
		S. A. Kivelson, 
		Breakdown of the Migdal-Eliashberg theory: A determinant quantum Monte Carlo study,
		Phys. Rev. B {\bf{97}}, 140501(R) (2018).
		%
		\bibitem{Esterlis19}
		I. Esterlis, S. A. Kivelson, and D. J. Scalapino,
		Pseudogap crossover in the electron-phonon system,
		Phys. Rev. B {\bf{99}}, 174516 (2019).
		%
		\bibitem{Chubukov20}
		A. V. Chubukov, A. Abanov, I. Esterlis, and S. A. Kivelson,
		Eliashberg theory of phonon-mediated superconductivity -- when it is valid and how it breaks down, Ann. Phys. {\bf{417}}, 168190 (2020).
		%
		\bibitem{Wang20}
		Y. Wang, I. Esterlis, T. Shi, J. I. Cirac, and E. Demler,
		Zero-temperature phases of the two-dimensional Hubbard-Holstein model: A non-Gaussian exact diagonalization study, Phys. Rev. Res. {\bf{2}}, 043258 (2020).
		%
		\bibitem{Sadovskii21}
		M. V. Sadovskii, Limits of Eliashberg Theory and Bounds for Superconducting Transition Temperature, arXiv: 2106.09948v1 [cond-mat.supr-con] 18 Jun 2021.
		%
		\bibitem{Cichutek22}
		N. Cichutek, M. Hansen, and P. Kopietz,
		Phonon renormalization and Pomeranchuk instability in the Holstein model,
		Phys. Rev. B {\bf{105}}, 205148 (2022).	
		%
		\bibitem{Fetter71}
		A. L. Fetter and J. D. Walecka, {\it{Quantum Theory of Many-Particle Systems}}, (McGraw-Hill, 
		New York, 1971).
		%
		\bibitem{Kopietz95}
		P. Kopietz, V. Meden, and K. Sch\"{o}nhammer, 
		Anomalous scaling and spin-charge separation in coupled chains,
		Phys. Rev. Lett. {\bf{74}}, 2997 (1995).
		%
		\bibitem{Kopietz97}
		P. Kopietz, V. Meden, and K. Sch\"{o}nhammer, 
		Crossover between Luttinger and Fermi liquid behavior in weakly coupled metallic chains,
		Phys. Rev. B {\bf{56}}, 7232 (1997).
		%
		\bibitem{Mahan10}
		G. D. Mahan, {\it{Many-Particle Physics}}, 3rd Edition, (Kluwer Academic/Plemum Publishers, New York, 2010).
		%
		\bibitem{Bohm51}
		D. Bohm and T. Staver, Application of Collective Treatment of Electron and Ion Vibrations to Theories of Conductivity and Superconductivity,
		Phys. Rev. {\bf{84}}, 836 (1951).
		%
		\bibitem{Ashcroft76}
		N. W. Ashcroft and N. D. Mermin, {\it{Solid State Physics}}, (Holt-Saunders, Philadelphia, 1976).
		%
		\bibitem{Grunerbook94}
		G. Gr\"{u}ener, {\it Density Waves in Solids}, (Perseus Pub. Cambridge, Massachusetts, 1994).
		%
		\bibitem{Kopietz96a}
		P. Kopietz, Bosonization of coupled electron-phonon systems, Z. Phys. B {\bf{100}},
		561 (1996).
		%
		\bibitem{Kopietz96b}
		P. Kopietz, {\it{Bosonization of Interacting Fermions in Arbitrary Dimensions}},
		(Springer, Berlin, 1996).
		%
		\bibitem{Mihaila11}
		B. Mihaila, Lindhard function of a $d$-dimensional Fermi gas, arXiv:1111.5337v1 [cond-mat.quant-gas] 2 Nov 2011.
		%
		\bibitem{Khomskii10}
		D. Khomskii, {\it{Basic Aspects of the Quantum Theory of Solids: Order and Elementary Excitations}}, (Cambridge University Press, Cambridge, 2010).
		%
		\bibitem{Girault89}
		S. Girault, A. H. Moudden, and J. P. Pouget,
		Critical x-ray scattering at the Peierls transition of the blue bronze,		Phys. Rev. B {\bf 39}, 4430 (1989).
		%
		\bibitem{footnoteGruener}
		The ratio $m^{\ast}/ m$ introduced in the review  by Gr\"{u}ner \cite{Gruener88} 
		is in our notation given by $m^*/m=1/g_1+1$ where the dimensionless coupling $g_1$ is defined in Eq.~(\ref{eq:g1defsingle}).
		Our result (\ref{eq:vpdef}) for the phason velocity of the Fr\"{o}hlich model agrees with the corresponding expression
		given by Gr\"{u}ner \cite{Gruener88} to leading order in $g_1$.
		%
		\bibitem{Giamarchi_Book} T. Giamarchi, {\it Quantum physics in one dimension} (Clarendon Press, Oxford, 2003).
		%
		\bibitem{Tsyplyatyev22}
		O. Tsyplyatyev,
		Splitting of the Fermi point of strongly interacting electrons in one dimension: A nonlinear effect of spin-charge separation,
		Phys. Rev. B {\bf 105}, L121112 (2022).
		\bibitem{Vianez22}
		P. M. T. Vianez, Y. Jin, M. Moreno, A. S. Anirban, A. Anthore, W. K. Tan, J. P. Griffiths, I. Farrer, D. A. Ritchie, A. J. Schofield, O. Tsyplyatyev, and C. J. B. Ford,
		Observing separate spin and charge Fermi seas in a strongly correlated one-dimensional conductor,
		Sci. Adv. {\bf 8}, 2781 (2022).
		%
		\bibitem{Fradkin83} E. Fradkin and J. E. Hirsch, Phase diagram of one-dimensional electron-phonon systems. I. The Su-Schrieffer-Heeger model,  Phys. Rev. B {\bf 27}, 1680 (1983).
		\bibitem{Weiss98} A. Wei\ss e and H. Fehske, Peierls instability and optical response in the one-dimensional half-filled Holstein model of spinless fermions, Phys. Rev. B {\bf 58}, 13526 (1998).
		\bibitem{Jackson75}
		See, for example, J. D. Jackson, {\it{Classical Electrodynamics}}, (2nd Edition, Wiley, New York, 1975), p. 285.
		%
		\bibitem{Ando05}
		H. Ando, T. Yokoya, K. Ishizaka, S. Tsuda, T. Kiss, S. Shin, T. Eguchi, M. Nohara, and H. Takagi,
		Angle-resolved photoemission study of K$_{0.3}$MoO$_3$: direct observation of temperature-dependent Fermi surface across the Peierls transition,		J. Phys.: Condens. Matter {\bf 17} 4935	(2005).	
		%
		\bibitem{Sato83}
		M. Sato, H. Fujishita and S. Hoshino,
		Neutron scattering study on the structural transition of
		quasi-one-dimensional conductor K$_{0.3}$MoO$_3$, 
		J. Phys. C: Solid State Phys. {\bf{16}}, L877 (1983).
		%
		\bibitem{Varlamov_book} A. Larkin and A. Varlamov, {\it Theory of Fluctuations in Superconductors}, (Oxford University Press, Oxford, 2005).
		%
		\bibitem{Degiorgi91}
		L. Degiorgi, B. Alavi, G. Mih\'aly, and G. Gr\"uner,
		Complete excitation spectrum of charge-density waves: Optical experiments on ${\mathrm{K}}_{0.3}$${\mathrm{MoO}}_{3}$,
		Phys. Rev. B {\bf 44},  7808 (1991).
		%
		\bibitem{Kurihara76}
		S. Kurihara,
		A Microscopic Theory of the Pinning Effect in Peierls Systems with Dilute Impurities,
		J. Phys. Soc. Jpn. {\bf 41}, 1488 (1976)		
		%
		\bibitem{Dzyaloshinskii74} I. E. Dzyaloshinskii and A. I. Larkin, Correlation functions for a one-dimensional Fermi system with long-range interaction (Tomonaga model), Sov. Phys. JETP {\bf 38}, 202 (1974).
		%
		\bibitem{Anderson70}
		P. W. Anderson, G. Yuval, and D. R. Hamann,
		Exact Results in the Kondo Problem. II. Scaling Theory, Qualitatively Correct Solution, and Some New Results on One-Dimensional Classical Statistical Models,
		Phys. Rev. B {\bf 1}, 4464 (1970).
		%
		\bibitem{Solyom79}
		J. Solyom,
		The Fermi gas model of one-dimensional conductors,
		Adv. Phys. {\bf 28}, 201 (1979).
		%
		\bibitem{Kurihara80}
		S. Kurihara,
		Nonlinear Amplitude-Phase Interaction in Charge-Density-Wave System,
		J. Phys. Soc. Jpn. {\bf 48}, 1821 (1980).
		%
		\bibitem{Takada85}
		S. Takada, K. Y. M. Wong, and T. Holstein,
		Damping of charge-density-wave motion,
		Phys. Rev. B {\bf 32}, 4639 (1985).
		%
		\bibitem{Hahn23}
		V. Hahn and P. Kopietz, Kinetic theory for collective modes in a charge-density wave, 
		(unpublished).		
		
		
	\end{thebibliography}
\end{document}